\documentclass[12pt,preprint]{aastex}
\usepackage{apjfonts}
\usepackage{amsmath}
\usepackage{amssymb}
\usepackage{lscape}
\usepackage{color}

\def \Se#1{Section~\ref{sec:#1}}

\def \Fig#1{Fig.~\ref{fig:#1}}
\def \Figs#1#2{Figs.~\ref{fig:#1}-\ref{fig:#2}}
\def \Figure#1{Figure~\ref{fig:#1}}
\def \Figures#1#2{Figures~\ref{fig:#1}-\ref{fig:#2}}

\def \Table#1{Table~\ref{tbl:#1}}

\def \eg{{e.g.~}}
\def \ie{{i.e.~}}

\def \arcsec{{$^{\prime\prime}$}}

\def \MLA{$M_*/L$}
\def \MLB{$M_*/L$~}
\def \MStrA{$M_*$}
\def \MStrB{$M_*$~}
\def \MLiA{$M_*/L_i$}
\def \MLiB{$M_*/L_i$~}
\def \MZerA{$M_*^{0D}$}
\def \MOneA{$M_*^{1D}$}
\def \MTwoA{$M_*^{2D}$}
\def \MZerB{$M_*^{0D}$~}
\def \MOneB{$M_*^{1D}$~}
\def \MTwoB{$M_*^{2D}$~}
\def \SigMA{$\Sigma_*$}
\def \SigMB{$\Sigma_*$~}
\def \tBurA{$t_{\text{burst}}$}
\def \tBurB{$t_{\text{burst}}$~}

\slugcomment{Accepted to MNRAS.} 

\shorttitle{Uncertainties of Stellar Mass Estimates}
\shortauthors{Roediger \& Courteau 2015}

\begin{document}


\title{On the Uncertainties of Stellar Mass Estimates via Colour Measurements}

\author{Joel C. Roediger$^{1,2,3}$ and St\'ephane Courteau$^1$}
\affil{$^1$ Department of Physics, Engineering Physics \& Astronomy, Queen's 
University, Kingston, Ontario, Canada}
\affil{$^2$ Department of Astronomy \& Astrophysics, University of California, 
Santa Cruz, CA, USA}
\affil{$^3$ National Research Council of Canada, Victoria, BC, Canada}

\email{Joel.Roediger@nrc-cnrc.gc.ca,courteau@astro.queensu.ca}


\begin{abstract}
Mass-to-light versus colour relations (MLCRs), derived from stellar population 
synthesis models, are widely used to estimate galaxy stellar masses (\MStrA) 
yet a detailed investigation of their inherent biases and limitations is still 
lacking.  We quantify several potential sources of uncertainty, using optical 
and near-infrared (NIR) photometry for a representative sample of nearby 
galaxies from the Virgo cluster.  Our method for combining multi-band 
photometry with MLCRs yields robust stellar masses, while errors in \MStrB 
decrease as more bands are simultaneously considered.  The prior assumptions in 
one's stellar population modelling dominate the error budget, creating a 
colour-dependent bias of up to 0.6 dex if NIR fluxes are used (0.3 dex
otherwise).  This matches the systematic errors associated with the method of 
spectral energy distribution (SED) fitting, indicating that MLCRs do not suffer 
from much additional bias.  Moreover, MLCRs and SED fitting yield similar 
degrees of random error ($\sim$0.1-0.14 dex) when applied to mock galaxies and, 
on average, equivalent masses for real galaxies with \MStrB $\sim$ 
10$^{8-11}$ M$_{\odot}$.  The use of integrated photometry introduces 
additional uncertainty in \MStrB measurements, at the level of 0.05-0.07 dex.  
We argue that using MLCRs, instead of time-consuming SED fits, is justified in 
cases with complex model parameter spaces (involving, for instance, 
multi-parameter star formation histories) and/or for large datasets.  
Spatially-resolved methods for measuring \MStrB should be applied for small 
sample sizes and/or when accuracies less than 0.1 dex are required.  An 
Appendix provides our MLCR transformations for ten colour permutations of the 
$grizH$ filter set.
\end{abstract}

\keywords{galaxies: clusters: individual: Virgo; galaxies: general; galaxies: formation; galaxies: fundamental parameters; galaxies: stellar content}


\section{Introduction}\label{sec:intro}

Stellar mass (\MStrA) is one of the most fundamental parameters for tracing the 
formation of galaxies, as embodied in its many correlations with other galaxy 
properties, such as halo mass, size, star formation rate, gas-phase and stellar 
metallicity, at all redshifts \citep{T04,G05,B10,M10}.  Knowledge of \MStrB 
also informs us on a variety of phenomena pertaining to galaxy formation 
efficiencies, dynamics, and feedback histories.  Given its astrophysical 
relevance, there is natural interest in measuring \MStrB as accurately as 
possible, whilst also achieving a complete account of its associated errors. 

For local galaxies, whose distances are fairly well-known, determining \MStrB 
essentially consists of measuring the mean mass-to-light ratio (\MLA) of their 
stellar populations.  The standard approach of measuring \MLB is to fit stellar 
population synthesis (SPS) models to the SED, line strength indices, and/or 
full spectrum of a galaxy.  Alternatively, one may access \MLB through 
maximal-disk fits \citep[e.g.][]{vAS86}, relations between \MLB and broadband 
colour \citep[e.g.][]{B03}, or a combination of the two \citep{BdJ01}.  These 
methods allow one to access \MLB whenever multi-band photometry or high-quality 
spectroscopy is unavailable.

In addition to minimal input, \MLA-colour relations (MLCRs) make determining 
mass-to-light ratios quite trivial, through the use of (quasi-)linear 
relations.  The linearity of MLCRs stems from the fact that changes in age, 
metallicity, and reddening scatter model stellar populations along the 
relations rather than away from them; equivalently, the 
age-metallicity-reddening degeneracy conspires to keep the scatter in MLCRs low 
\citep{BdJ01}.  The simplicity of MLCRs has precipitated their widespread use, 
as well as several calibrations of these relations 
\citep{BdJ01,B03,Z09b,GB09,T11,IP13,MS14} [hereafter BdJ01, B03, GB09, Z09, 
T11, IP13, and MS14 respectively].  These calibrations differ largely by virtue 
of their underlying SPS treatments, including the effects of dust attenuation, 
bursts of star formation and chemical evolution.  These effects can also be 
modelled via other methods for measuring \MLB (\eg SED fitting), though usually 
at the cost of increased computational expense.  The MLCR method suffers no 
such penalty for added model complexity though, making it an expedient, and 
thus practical (if reliable), route to \MLB in most circumstances.

Despite their widespread use, an exhaustive analysis of the accuracy of 
MLCRs is still lacking (GB09 being the most in depth thus far).  BdJ01 showed 
that uncertainties in their adopted treatments of SPS, galaxy evolution, and 
dust translated into a wavelength-dependent error of 0.1-0.2 dex in \MLA while 
large bursts of recent star formation (10\% by mass) inflate this range to 
0.3-0.5 dex.  Although B03 echoed these conclusions, the BdJ01 MLCRs targetted 
spiral galaxies alone (as did MS14) and the B03 MLCRs excluded treatments for 
reddening and bursts of star formation.  Z09 argued that omitting the latter 
effects results in systematically larger \MLB values at blue colours and 
smaller values at red colours.  Using mock galaxies, GB09 studied a multitude 
of effects on the recovery of \MLA, finding, amongst other things, that with 
sufficient signal-to-noise ($S/N$; $>$30), individual optical colours performed 
as well as spectroscopic indices in the median, except for galaxies described 
intermediate-age populations and a bursty star formation history (SFH).

To be sure, effects other than our assumptions about the SFHs and dust contents 
of nearby galaxies contribute to the uncertainty budget underlying MLCRs.  Z09 
compared MLCRs derived from the \citet[hereafter BC03]{BC03} SPS model and the 
unpublished 2007 update thereof (hereafter CB07), where the greatest difference 
between the two presumably lay in their prescriptions for TP-AGB stars.  They 
demonstrated that this phase of stellar evolution alone can alter \MLB 
predictions by up to 0.1 and 0.4 dex at optical and near-infrared (NIR) 
wavelengths, respectively.  T11 partly explored the effect of constraining 
MLCRs with SED fits to real galaxies.  While their empirical MLCR differs 
markedly in shape from the one defined by their prior-weighted SPS model 
library (see their Fig. 13), the two appear to be statistically consistent 
everywhere except in the interval $g-i$ = 0.7-1.05.  MS14 presented corrections 
to several sets of published MLCRs (including B03, Z09, and IP13) for 
inconsistencies in predicted \MLB throughout the optical-NIR wavelength range, 
presumably tied to uncertainties in stellar evolution, although the corrections 
were solely based on a limited sample of nearby spiral galaxies and the degree 
of inconsistency was not quantitatively established.  Finally, the choice of 
IMF is a crucial parameter of MLCRs as it controls their normalisations 
(zero-points).  For example, \MStrB estimates based on a Chabrier or Salpeter 
IMF differ by 0.3 dex, with the latter being higher.

Most, if not all, of the systematics inherent to the MLCRs have also been 
studied in the context of SED fitting, all yielding similar results.  While 
much effort has been spent on understanding the latter (see \citealt{C13} and 
\citealt{C14} for recent reviews), the computational expense associated with 
even modest parameter spaces severely complicates the investigation of certain 
issues via SED fitting.  For instance, masses derived from integrated 
photometry are biased to the brightest regions within galaxies (\ie 
light-weighted) and therefore susceptible to erroneous \MLB determinations.  
Two clear causes of this are interstellar dust or bright stars, which can 
shroud or outshine a potentially large fraction of a galaxy's stellar mass.  
This bias can be overcome by adding up the mass in galaxies pixel-by-pixel.  
Given the many pixels that local galaxies fill with modern imaging cameras, 
measuring \MStrB pixel-by-pixel through SED fits is potentially expensive.  
Thus in their study of spatially-resolved stellar masses, Z09 drew on the MLCR 
method.  They found that the pixel-by-pixel approach yields masses that are 
larger than their integrated counterparts by up to 0.2 dex, depending on the 
filters and SPS models used.  The Z09 sample however was small (9 galaxies) and 
skewed to late-type spirals (5/9 had Sbc/Sc morphologies).  An expanded sample 
would cement their findings on firmer grounds\footnote{During the final review 
of this paper, the manuscript by \cite[hereafter SS15]{SS15} expanded upon 
Z09's findings by measuring pixel-by-pixel masses for 67 nearby galaxies 
comprising 9 E/S0's, 4 Irr's and the rest spirals.  SS15 stated that the 
discrepancy between resolved and integrated masses grows with higher specific 
star formation rates and disappears for spatial resolutions $\ge$3 kpc.}

We wish to address several open issues pertaining to MLCRs for the measurement 
of \MStrA.  While earlier studies have focussed on comparisons of theoretical 
MLCRs against one another for various combinations of colours and $M_*/L_X$ 
(the mass-to-light ratio in filter $X$), we also take the approach of comparing 
them against {\it real} galaxies.  To this end, we conduct several tests of 
MLCRs using a representative sample of local galaxies drawn from the ``SHIVir 
survey'' of Virgo Cluster galaxies, for which deep luminosity profiles at 
optical and NIR wavelengths are available \citep{M11}.  We also consider {\it 
multiple} colours at once in fitting for \MLA, whereas previous analyses used 
two colours at most (Z09).  We specifically address \MStrB differences due to: 
(i) MLCRs versus SED fitting; (ii) different MLCR sets; (iii) limited SED 
constraints; and (iv) integrated versus spatially-resolved photometric data.

Our paper is organized as follows.  In \Se{data}, we introduce the SHIVir 
survey upon which our stellar masses are derived.  In \Se{model}, we construct 
our own MLCRs, including a discussion of our assumed priors, and test the 
fidelity of SED-based masses using mock galaxies, which allows us to rank the 
systematics inherent to the MLCR method.  The latter are investigated and 
quantified (using real galaxies) in \Se{r&d}.  Finally, our conclusions and 
suggestions for future work are reported in \Se{conc}.


\section{Data}\label{sec:data}

Thanks to its proximity and richness, the Virgo Cluster enables the 
simultaneous analysis of the stellar populations in all types of present-day 
galaxies on both spatially-resolved and integrated scales.  The SHIVir survey 
("{\bf S}pectroscopy and {\bf $H$}$-$band {\bf I}maging of {\bf Vir}go cluster 
galaxies"; \citealt{M11}) seizes this opportunity by combining optical and NIR 
imaging for a volume-limited sample of Virgo galaxies with apparent $B-$band 
magnitudes $>$ 16 and located $\lesssim$1.7 Mpc (in projection) from the core 
of the sub-cluster A, taken to be M87.  Like the cluster itself, the sample 
includes the full range of morphological types and is thus representative of 
the low-redshift galaxy population.  For these reasons, we draw on the SHIVir 
survey for our tests of the MLCR method.

The SHIVir galaxies were imaged in four optical filters ($griz$) and the NIR 
$H-$band.  The optical imaging was downloaded from the seventh data release of 
the Sloan Digital Sky Survey \citep[SDSS;][]{A09}.  Conversely, the NIR imaging 
is a heterogeneous collection obtained from either the GOLDMine database 
\citep{G03}, Two Micron All Sky Survey \citep[2MASS;][]{S06}, or an extensive 
observational campaign conducted by our team with Mauna Kea telescopes and 
detectors.  SHIVir data products include luminosity profiles, integrated 
magnitudes, effective radii and surface brightnesses, and concentrations for 
286 Virgo cluster galaxies in the $grizH$ filters.  Further details about 
SHIVir photometry and its associated data products can be found in 
\cite{M09,M11} and at this website\footnote{See also {\tt 
http://www.astro.queensu.ca/virgo}.}.


\section{Modelling the Stellar Mass-to-Light Ratio}\label{sec:model}

We present our adopted method for modelling the stellar mass-to-light ratios of 
SHIVir galaxies (\Se{mod-m2l}).  Recall that a major aim of this work is to 
assess whether \MStrB derived from integrated photometry are consistent with 
those based on spatially-resolved data.   Since the latter typically involve 
modelling \MLB for thousands of pixels per galaxy, our method invariably draws 
upon MLCRs for their expediency.  Although several calibrations of MLCRs 
already exist in the literature (B03, Z09, T11, IP13), our experiment calls for 
the development of our own MLCRs, as argued below (\Se{mod-m2l}).  We also 
develop in \Se{mod-mock} mock galaxy tests for the recovery of \MStrB based on 
SED fitting.  Finally, in \Se{mod-p&m} we outline our approach to measuring 
\MStrB through 1D light profiles and 2D maps of the stellar mass density 
(\SigMA) distribution in SHIVir galaxies.

\subsection{Mock Galaxy Tests of SED Fitting}\label{sec:mod-mock}

This work focusses, in part, on modelling \MStrB for a representative sample of 
local galaxies whose light (and thus mass) can be measured in three ways that 
we refer to as 0D, 1D, and 2D.  `0D' refers to the end point of a galaxy curve 
of growth, \ie the total luminosity of the galaxy up to a certain radius.  `1D' 
refers to the profile of galaxy surface brightness as a function of 
galactocentric radius; the brightness at each radius represents the median flux 
divided by the area subtended by the associated isophote.  Lastly, `2D' stands 
for the 2-dimensional image of a galaxy whose pixels reveal information about 
the local stellar population.

The principal unknowns governing \MLB for any stellar population are its IMF, 
age, metallicity ($Z$), and reddening.  Given the wide range of scales explored 
in this work, our modelling must therefore include populations covering a 
variety of evolutionary states.  Specifically, we require a library of 
theoretical stellar populations spanning extensive ranges in star formation 
rates and histories, chemistries, and dust contents.

The MAGPHYS library \citep[{\bf M}ulti-wavelength {\bf A}nalysis of {\bf 
G}alaxy {\bf PHYS}ical Properties;][]{dC08} is well-suited for our purposes.  
This library consists of 50,000 randomly-generated stellar and dust emission 
SEDs that can be self-consistently linked to interpret UV through sub-mm 
imaging of galaxies in terms of their stellar population and ISM properties.  
We disregard the dust emission SEDs since constraining ISM properties is not 
our current concern.  Assuming that dust emits at wavelengths $\ge$2.5 $\mu$m, 
we use only UV-NIR SED information from MAGPHYS.  This wavelength range 
comfortably captures the baseline spanned by the SHIVir database.

The MAGPHYS stellar SEDs can be computed based on either the BC03 or CB07 SPS 
models.  These SEDs are computed according to SFHs prescribed by five 
parameters.  These are the start time (1) and exponential decay rate (2) of 
continuous star formation, as well as the ages (3), durations (4), and mass 
fractions (5) of superimposed bursts, if any.  No prior constraints are made as 
to the number of bursts that underlie any given SED other than that half of 
them reflect a burst that occurred within the past 2 Gyr.

Attenuation is folded into the stellar SEDs following the \cite{CF00} 
two-component model for dust in molecular clouds and the ambient ISM.  The 
amount of attenuation suffered by each SED is described by two parameters: the 
total optical depth towards stars $<$0.01 Gyr-old and the fractional 
contribution of ambient dust to that optical depth ($\mu$).  The SED 
metallicities are also allowed to cover the range $Z$ = 0.0004-0.04.(\ie 0.02-2 
$Z_{\odot}$).  Altogether, eight parameters are used to describe the stellar 
SEDs.  To compute each SED, the MAGPHYS software randomly draws values of these 
parameters from prior distributions.  See \cite{dC08} for more information 
about these priors.

The extensive parameter space covered by the MAGPHYS library encapsulates the 
variety of stellar populations within present-day galaxies.  Accordingly, this 
library may be used to evaluate the accuracy of \MStrB measurements from SED 
fitting.  Specifically, we wish to know the effects that $S/N$ and SED sampling 
have on these measurements.  Using mock datasets from MAGPHYS, \cite{dC08} were 
able to recover input stellar masses with very high fidelity in the case of an 
exquisitely sampled SED (\ie UV through sub-mm fluxes) and a fixed uncertainty 
of 10\% in each filter.  The authors also claimed that the masses measured from 
the integrated SEDs of three galaxies (NGC 0337, 3521; Mrk 33) remained 
relatively robust even as UV, Balmer, far-IR and sub-mm fluxes were iteratively 
excluded from their fits.  These results, however, were not quantified.  
Moreover, the authors did not quantify the variations of their \MStrB estimates 
if their only available data consisted of optical plus NIR fluxes, a situation 
that we face here and often encountered in the literature.

To assess the quality of stellar masses that can be achieved with optical and 
NIR photometry, we have conducted our own mock galaxy tests for various 
combinations of $S/N$ and SED sampling.  For each test, 500 randomly selected 
MAGPHYS stellar SEDs had Gaussian noise of a fixed percentage applied to their 
fluxes from a select filter suite.  Stellar masses were then measured from 
these perturbed SEDs using the Bayesian fitting software included in the 
MAGPHYS package.

\begin{figure*}
 \begin{center}
  \includegraphics[width=0.9\textwidth,trim=0 150 0 150]{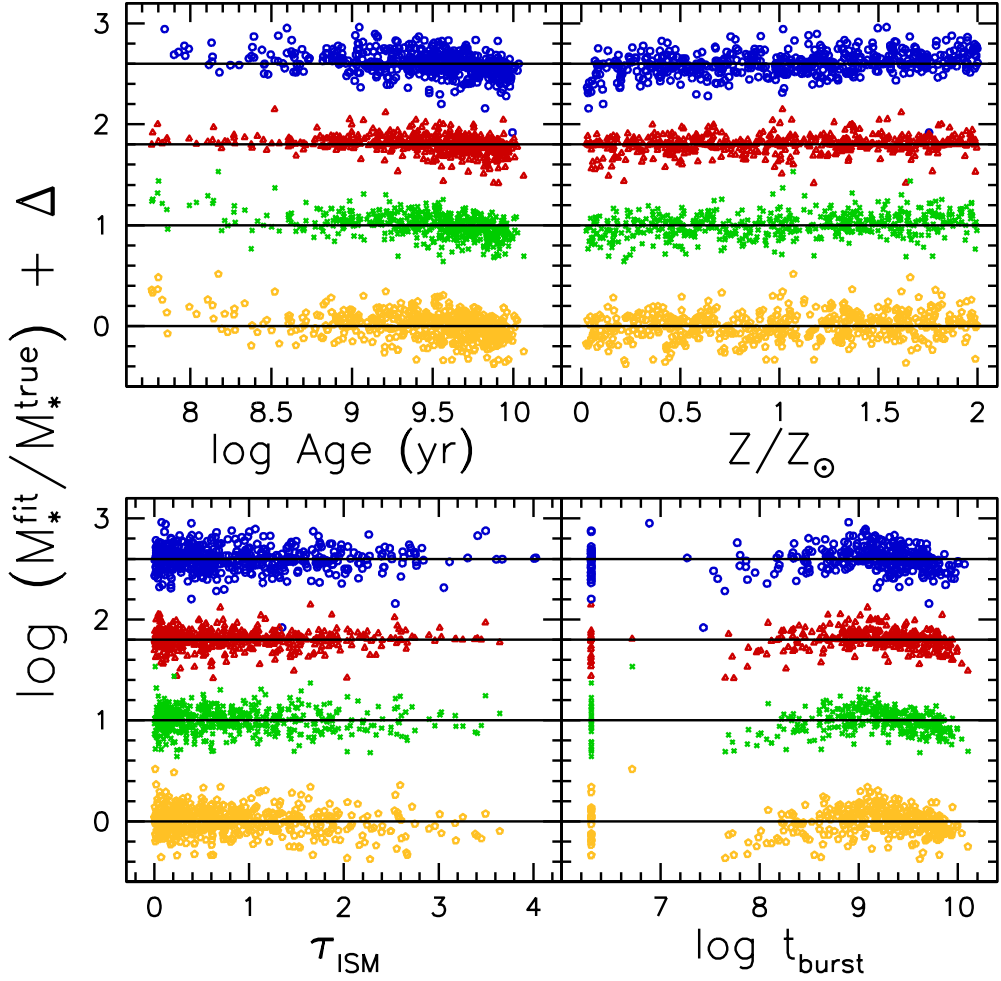}
  \caption{Recovery of the known stellar masses (\MStrA) of 500 mock galaxies 
  via Bayesian SED modeling [$M^{\text{fit}}_*$ = fitted \MStrA; 
  $M^{\text{true}}_*$ = true \MStrA], plotted as a function of mass-weighted 
  age, total metallicity ($Z/Z_{\odot}$), ISM optical depth 
  ($\tau_{\text{ISM}}$) and age of the latest burst of star formation 
  (\tBurA).  The galaxies were drawn from and fitted with the MAGPHYS library, 
  based on the \citet[hereafter BC03]{BC03} stellar population synthesis (SPS) 
  model.  The experiment is carried out for four combinations of SED sampling 
  and signal-to-noise ratio ($S/N$): (a) UV to near-infrared coverage and $S/N$ 
  = 10 (blue points); (b) as in (a) but $S/N$ = 20 (red); (c) SDSS $griz$ + 
  2MASS $H$, $S/N$ = 20 (green); and (d) SDSS $gi$ (orange), $S/N$ = 20.  The 
  filter combinations examined in cases (c-d) focus on the data available from 
  the SHIVir survey \citep{M11}.  Datasets (a-c) have been offset vertically 
  for display purposes (denoted by the solid lines).  Galaxies which have 
  experienced a burst during the most recent timestep (\ie \tBurB = 0.0) have 
  been assigned to log \tBurB = 6.3, also for display purposes.  The residual 
  distribution is similar in all cases, having a mean of roughly zero and 
  dispersion of $\sim$0.1 dex.}
  \label{fig:sed-vs-true}
 \end{center}
\end{figure*}

\Figure{sed-vs-true} demonstrates the accuracy to which MAGPHYS can 
recover \MStrB for galaxies and how that accuracy might vary with galaxy 
parameters.  The parameters studied include mass-weighted age, metallicity,
ISM optical depth and age of the most recent burst of star formation (\tBurA).  
Results are shown for four cases: (i) SEDs perturbed by 10\% noise and sampled 
with the GALEX $FUV$ and $NUV$, SDSS $ugriz$, and 2MASS $H$ filters [blue 
points]; (ii) as in (i) but with 5\% noise [red]; and (iii) as in (ii) but with 
SED sampling restricted to the $grizH$ filters [green]; or (iv) using $gi$ 
filters only [orange].  Cases (i) and (ii) in \Fig{sed-vs-true} reflect the 
situation of a well-sampled stellar SED while cases (iii) and (iv) match the 
data available from the SHIVir survey.  Case (iv) examines the impact of 
removing $r-$, $z-$, and $H-$band data, on account of the nebular 
contamination, low $S/N$, and/or poor sky subtraction that sometimes 
compromises galaxy photometry at those wavelengths.

In all four cases, the masses recovered by SED fitting are described by a 
residual distribution having zero mean and dispersion of 0.09-0.13 dex, \ie 
random errors of 23-30\%.  The robustness of the residuals in \Fig{sed-vs-true} 
is remarkable given the wide range in data richness explored.  We expect that 
the constraints from well-sampled SEDs should better differentiate themselves 
from those of coarser SEDs as the $S/N$ increases beyond what we have 
explored.  Furthermore, accuracies of $\sim$0.1 dex can be achieved with the 
SHIVir filters and $S/N$ = 20.  Such a threshold $S/N$ translates into flux and 
colour errors of $\sim$0.05 mag and $\sim$0.10 mag, respectively.  The latter 
error will guide the construction of our own MLCRs, described in the following 
section.

\Fig{sed-vs-true} also hints at a possible, but complex, trend in accuracy with 
burst age, whereby masses are maximally overestimated around \tBurB $\sim$ 1-2 
Gyr and underestimated at the youngest and oldest \tBurB (exclusive of 
\tBurB).  This trend is consistent with the results of GB09, who found that, 
for an error of 0.1 mag in $g-i$, the \MLB of mock galaxies described by bursty 
SFHs and young/intermediate-age populations have predicted masses with median 
residuals of 0.0/0.15 dex.  It is worth noting that the trend appears weaker 
for the red points, which have higher $S/N$.  These results suggest that bursts 
are poorly constrained with broadband photometry of marginal quality ($S/N \le$ 
10), no matter how many filters one has data for.  Further work aimed at 
addressing the combination of filters (including narrow bands) and $S/N$ 
required to unravel the contribution of bursts to the recent SFHs of galaxies 
would be useful.

\subsection{Estimating \MLB via Colour Information}\label{sec:mod-m2l}

In \Se{mod-mock}, we quantified the internal accuracy of \MStrB measurements 
from SED fitting.  This provides us with a benchmark to evaluate the relative 
success of our adopted method (MLCRs) for measuring \MLB (\Se{r&d-mlcr}).
As stated earlier, the appeal of MLCRs lies in their minimal data requirements
and expediency in all applications.  

Since the pioneering work of BdJ01 and B03, MLCRs have been a popular tool for 
measuring \MLA.  Similar to the strategies of Z09, GB09, and MS14, we do not 
restrict our modelling to a single colour.  The motivation for this stems from 
the fact that every colour has a unique pattern of sensitivities to the 
parameters that determine \MLA.  As a result, the MLCR for any colour contains 
scatter (T11), which should be accounted for by modelling multiple colours at 
once.  However, using all of the $N$($N$-1)/2 colours that may be measured from 
an SED sampled with $N$ filters would implicitly weigh certain parts of the SED 
more than others.  Moreover, modelling all possible colours at once is 
tantamount to overfitting the data.  For any given SED then, we only consider 
as many colours as are necessary, when paired with the flux in a single band, 
to reconstruct the original SED\footnote{The fact that a flux is required to 
turn \MLB into \MStrB means that the number of colours needed to reconstruct a 
given SED is minimized.  For instance, a $griH$ SED is fully described by the 
$i-$band flux plus the colours $g-r$, $g-i$, and $g-H$.}.  We select those 
colours whose MLCRs have the least amount of scatter; that is, those which are 
in principle the strongest discriminants of \MLA.  The fact that we can model 
several colours at once makes our approach unique from those of Z09, GB09, and 
MS14, each of which considered two colours at most.

\begin{figure*}
 \begin{center}
  \includegraphics[width=0.9\textwidth]{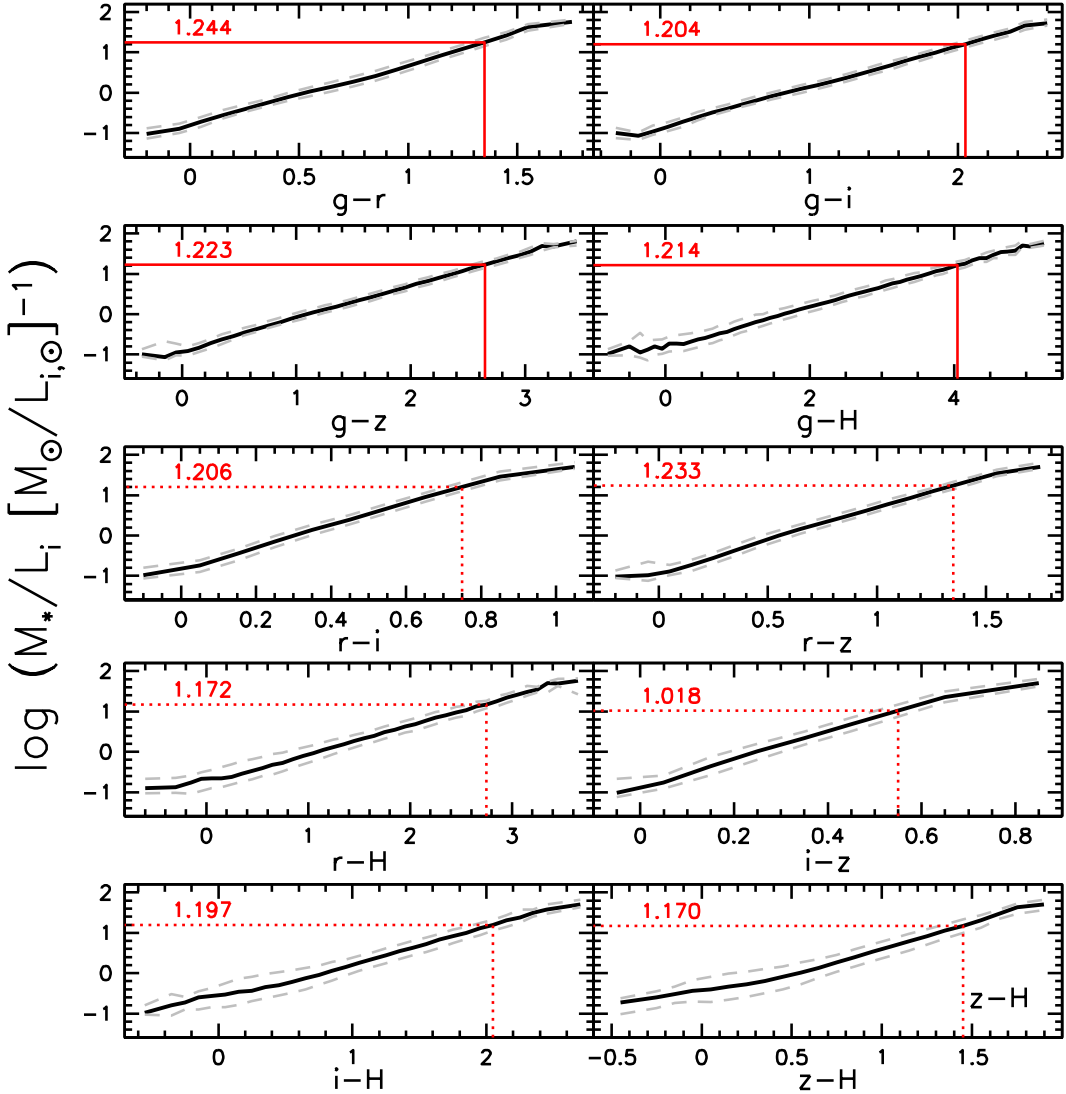}
  \caption{Demonstration of the simultaneous use of multiple colours to 
  constrain the stellar mass-to-light ratio (\MLA) via \MLA$-$colour relations 
  (MLCRs; constructed using the MAGPHYS/BC03 library).  The black line in each 
  panel traces the median \MLiB as a function of the particular colour plotted 
  along the $x-$axis, while the gray lines indicate the 16$^{\text{th}}$ and 
  84$^{\text{th}}$ percentiles.  Here we focus on the colours that can be 
  measured for galaxies with the SHIVir dataset.  Red lines mark the \MLiB 
  values that would be measured for a hypothetical population, given its 
  integrated $grizH$ SED.  Since four colours plus the $i-$band flux is the 
  minimum amount of information required to uniquely describe such an SED, we 
  differentiate the red lines into solid/dotted types according to those 
  colours that provide the strongest/weakest constraints on \MLiA.  Our 
  procedure then combines the strongest estimates via a weighted average to 
  obtain the overall \MLiB for the population in question.  While this demo is 
  based on integrated data, it is straightforward to generalize our procedure 
  to work with colour profiles/maps instead.}
  \label{fig:mlcr-demo}
 \end{center}
\end{figure*}

\Figure{mlcr-demo} highlights our multi-MLCR method for measuring \MLB.  We 
show with black lines theoretical trends in median \MLiB versus all ten colours 
that can be measured from the full SHIVir SED.  These trends were determined 
from the BC03 version of the MAGPHYS library.  The grey lines trace the 
16$^{\text{th}}$ and 84$^{\text{th}}$ percentiles of the model distribution for 
each MLCR, which we take to be the 1-$\sigma$ uncertainty in \MLiA.  The red 
lines indicate the \MLiA values that we would infer for a random MAGPHYS model 
according to its integrated $grizH$ SED.  The MLCRs for $g-r$, $g-i$, $g-z$, 
and $g-H$ provide the tightest constraints on \MLiB for an SED sampled with 
those filters, and so are the ones that we would use in this case; hence the 
solid red lines in those panels while all the others are dotted.  Comparisons 
of the \MLiB values inferred for SHIVir galaxies based on these colours 
confirms that, on average, they are mutually consistent within 1$\sigma$.  
Thus by comparing a set of measured colours -- be it for an individual pixel, 
isophote, or whole galaxy -- to such MLCRs as displayed in \Fig{mlcr-demo}, we 
can deduce a corresponding set of \MLiB measurements.  These in turn may be 
combined to obtain a single \MLiB value for the object in question.  We use a 
weighted average, based on the 16-84$^{\text{th}}$ percentile range (linear 
scale), in order to reduce the contribution from colours that are poorer \MLiB 
discriminants.  This is how we use multiple MLCRs simultaneously to measure 
\MLA.

\begin{figure*}
 \begin{center}
  \includegraphics[width=0.9\textwidth]{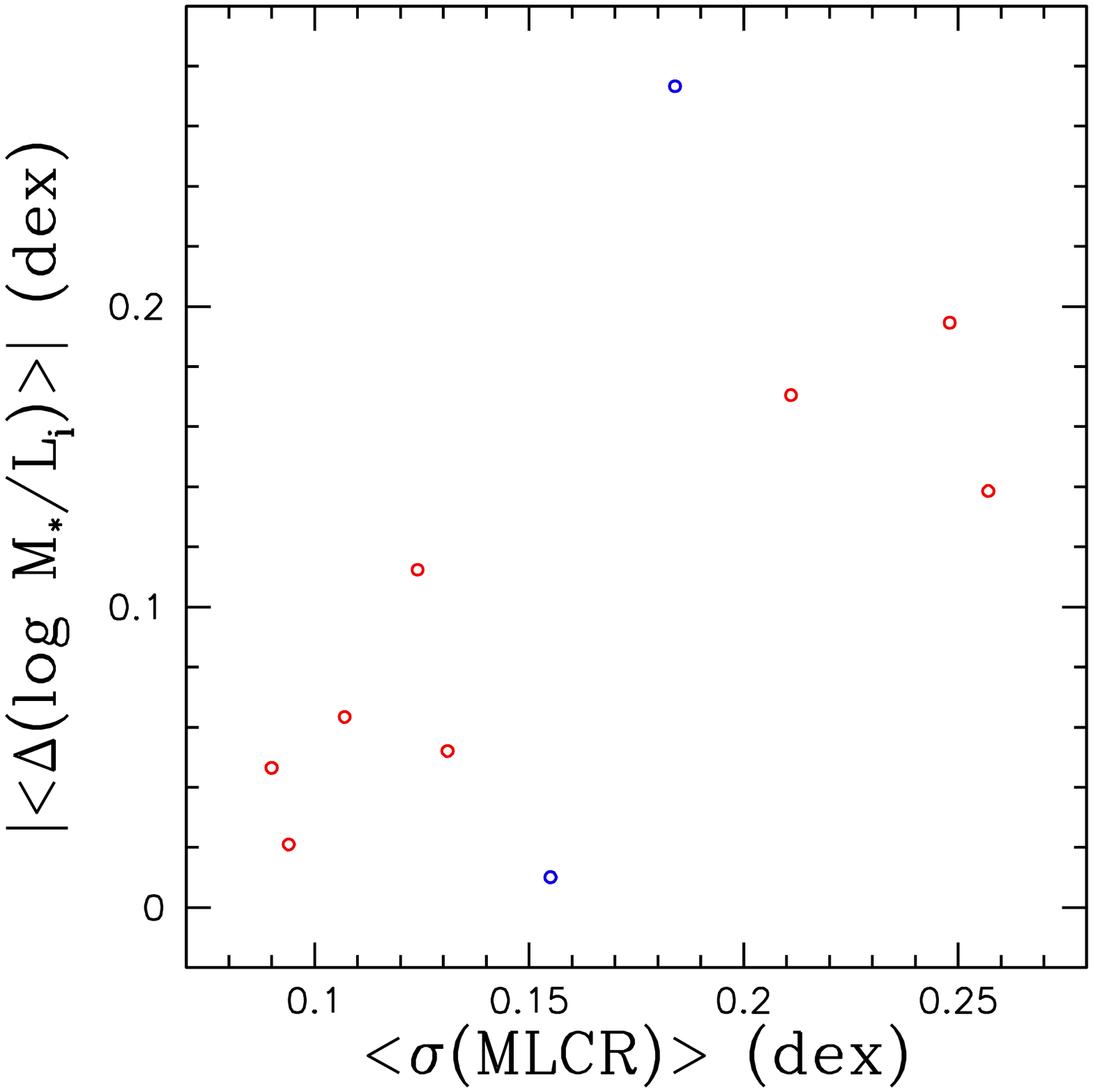}
  \caption{Absolute mean offset of \MLiB predicted by MLCRs versus the mean 
  scatter in the relations.  Each data point correspnds to a colour that can be 
  accessed by the SHIVir survey (\eg $g-i$) and the value along the vertical 
  axis represents an average over the 280 galaxies contained therein.  Offsets 
  were measured with respect to the \MLiB determined by fitting the $grizH$ 
  SEDs of the SHIVir galaxies directly.  Aside from a couple outliers, there is 
  a clear correlation between the two variables, implying that the precision of 
  a given MLCR is related to its accuracy.}
  \label{fig:dm2l-vs-dmlcr}
 \end{center}
\end{figure*}

A potential pitfall of our method is that, in trying to minimize the random 
error, we increase the systematic error above the level incurred were we to use 
MLCRs with larger scatter instead.  To show that this is in fact not the case, 
we plot in \Figure{dm2l-vs-dmlcr} the mean offset of the MLCR-based \MLiB 
values predicted from each of the 10 colours accessed by the SHIVir dataset as 
a function of the mean scatter in the associated MAGPHYS/BC03 MLCRs.  The 
offsets are computed with respect to the \MLiB inferred from fitting the 
$grizH$ SEDs with the same model library.  The figure convincingly demonstrates 
that, despite a couple of outliers, there is a clear correlation between the 
two variables.  In other words, the precision of a given MLCR is, on average, 
directly related to the accuracy of its predictions and our method is 
physically sound.  We have also investigated the impact of folding systematic 
errors though our mock galaxy tests (\Fig{sed-vs-true}) into our weighting 
scheme and found them to be minimal.

To apply our method to SHIVir data, we require MLCRs cast in terms of both the 
SDSS and 2MASS/$H$ bands.  Existing MLCRs prove inadequate for our purposes 
because they either do not treat both optical and NIR colours (\eg Z09) or they 
are based on simplistic priors (\eg B03).  We also wish to measure \MStrB 
through \SigMB profiles and maps, in which case our MLCRs must incorporate SFHs 
that are generally not smooth.  Finally, our desire to encompass all 
morphologies, as in SHIVir, favors MLCRs that span a wide variety of stellar 
populations.  For these reasons, we derived our own MLCRs, based on the MAGPHYS 
stellar SEDs.

We calibrate our MLCRs in terms of \MLiB and all possible colour combinations 
formed from the $grizH$ filters.  Although NIR wavelengths are often viewed as 
yielding the most robust \MLB values (\eg BdJ01), the current uncertainties in 
NIR fluxes predicted for TP-AGB stars make us favour an optical band instead 
\citep{M05,B07}.  For each relevant colour, we calibrate the associated MLCR 
with the MAGPHYS/BC03 library in the following way.  First, \MLiB and the 
colour of interest are computed for each SED in the library, where the former 
is based on the present-day \MStrA.  Those SEDs are then grouped by this colour 
into 0.1 mag-wide bins.  The median plus the 16$^{\text{th}}$ and 
84$^{\text{th}}$ percentiles of the \MLiB distribution in each bin are then 
calculated, where the percentiles are used to approximate the $\pm$1-$\sigma$ 
uncertainties.  This straightforward procedure yields the desired MLCR.  Note 
that the choice of bin width was made based on \Fig{sed-vs-true}, which shows 
that optical-to-NIR SEDs having $S/N$ = 20 yield stellar masses of comparable 
accuracy to those with better sampling.  Furthermore, surface brightness 
measurements with $S/N \ge$ 20 may be achieved in the outskirts of galaxies 
through, e.g., isophotal fitting.  For colours having errors $>$0.1 mag, we 
adjust the 1-$\sigma$ uncertainties on their \MLiB measurements to reflect the 
16$^{\text{th}}$ and 84$^{\text{th}}$ percentiles of the bins in which their 
lower and upper limits fall, respectively.  Although our approach to such cases 
favours larger uncertainties, a more accurate treatment would introduce 
precisely the type of computational intensity we wish to avoid with our method.

\subsection{Stellar Mass Density Profiles and Maps}\label{sec:mod-p&m}

\Fig{mlcr-demo} demonstrates how MLCRs reduce the measurement of \MStrB to a 
trivial algorithm.  This defining aspect of the MLCR method makes it highly 
desirable when considering, say, SED modelling for hundreds of thousands of 
resolution elements per galaxy.  Recall that the motivation to compute \MStrB 
from 2D maps stems from Z09's finding (corroborated by SS15) that the mass 
assessment is more accurate that way.

We now describe the additional preparations to infer \MStrB for galaxies based 
on either multi-band luminosity profiles or full images.  Note that we shall 
refer to masses derived from integrated luminosities, surface brightness 
profiles, and images as \MZerA, \MOneA, and \MTwoA, respectively.

\begin{figure*}
 \begin{center}
  \includegraphics[width=0.9\textwidth]{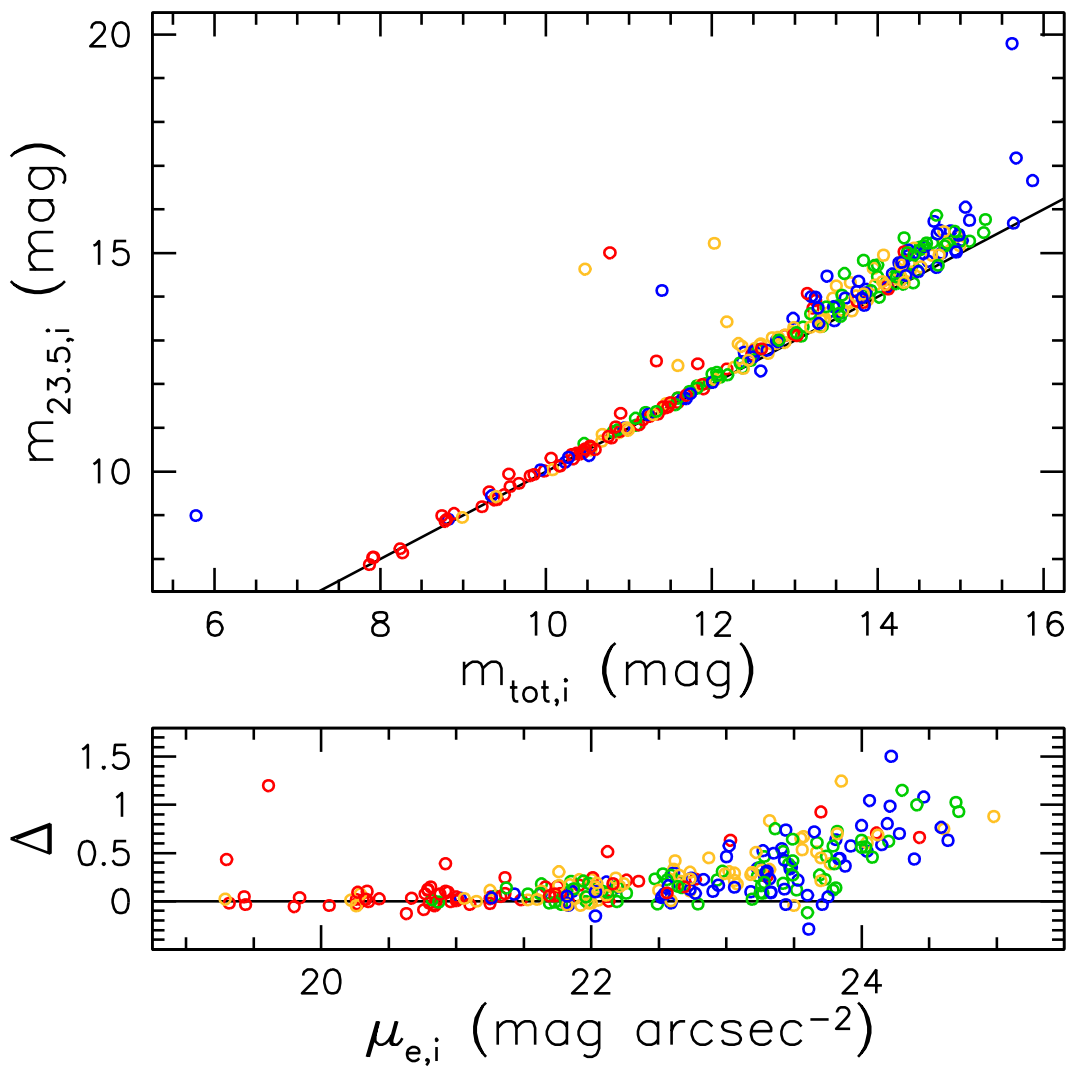}
  \caption{(top) Comparison of $i-$band aperture and extrapolated magnitudes 
  ($m_{23.5,i}$ and $m_{tot,i}$) for SHIVir galaxies, where the former have 
  been measured at the 23.5 mag arcsec$^{-2}$ isophotal radius.  The points are 
  colour-coded according to galaxy concentration, $C_{28}$ [$C_{28} <$ 2 .86 
  (blue), 2.86 $\ge C_{28} <$ 3.31 (green), 3.31 $\ge C_{28} <$ 3.99 (orange), 
  $C_{28} \ge$ 3.99 (red)] and the solid line marks the 1:1 relation.  (bottom) 
  The residuals between these quantities ($\Delta = m_{23.5,i} - m_{tot,i}$) 
  are plotted as a function of $i-$band surface brightness.  Not surprisingly, 
  $m_{23.5,i}$ breaks down as a proxy for total light at $\mu_{e,i} \ge$ 23.5 
  mag arcsec$^{-2}$; our conclusions are not affected by this caveat though.}
  \label{fig:m23d5-vs-mtot}
 \end{center}
\end{figure*}

For the purpose of comparing these various sets of masses, it is crucial that 
they be measured at a common galactocentric radius.  We choose the $i-$band 
23.5 mag arcsec$^{-2}$ isophotal radius ($R_{23.5,i}$) since it and the 
enclosed light ($m_{23.5,i}$) yield the tightest scaling relations \citep{H12} 
and are good proxys for the total sizes and luminosities of galaxies of 
intermediate-to-high surface brightnesses.  This is demonstrated in the top 
panel of \Figure{m23d5-vs-mtot}, which plots $m_{23.5,i}$ against the 
extrapolated magnitude in the same band ($m_{tot,i}$) for the SHIVir galaxies.  
The former were extracted from the SHIVir surface brightness profiles, while 
the latter were measured by \cite{M11}.  The points have been colour-coded by 
galaxy concentration and the solid black lines mark equality.  Three-quarters 
of the SHIVir sample have $(m_{23.5,i}-m_{tot,i}) <$ 0.3 mag, defined by high 
and intermediate surface brightness galaxies, while departures from $m_{23.5,i} 
\sim m_{tot,i}$ grow towards lower surface brightnesses, as shown in the bottom 
panel of \Fig{m23d5-vs-mtot}.  Since our main concern in this work is whether 
multiple colour measurements can be used to reliably constrain \MLA, these 
departures should not affect any of our conclusions.

\subsubsection{\MZerB \& \MOneA}\label{sec:mod-0+1d}

Since our methods for measuring \MZerB and \MOneB of a galaxy are intertwined, 
we shall discuss both herein.  Both approaches begin with the $grizH$ light 
profiles from SHIVir.  Depending on the bands being modelled, we interpolate 
the profiles (locally) for each galaxy to the largest pixel scale therein.  No 
interpolation is required when modelling optical SEDs alone, since the original 
data come from a single source (SDSS; pixel scale = 0.396\arcsec).  On the 
other hand, the heterogeneous nature of the NIR imaging -- having pixel scales 
in the range = 0.25-1.61\arcsec -- imply that either those or the optical 
profiles must be interpolated to a common scale when optical-NIR SEDs are 
considered.  Interpolated quantities include the surface brightness, 
ellipticity, and flux enclosed within each elliptical isophote.  We then 
correct for Galactic reddening using the updated version of the \cite{S98} 
prescription by \cite{SF11}.

In order to match the resolution of our MLCRs (0.1 mag; \Se{mod-m2l}), we use 
only light profiles with $S/N \ge$ 20 per isophote, per filter.  To meet this 
requirement, we bin the isophotes for all of the profiles of a given galaxy, 
marching outwards from the center. The bin sizes are increased until the $S/N$ 
threshold is met in all bands.  This procedure requires as input the $S/N$ as a 
function of galactocentric radius for each profile.  To conform with the 
adaptive smoothing technique that we apply in our 2D approach (\Se{mod-2d}), we 
estimate the $S/N$ of each isophote assuming background-dominated noise.  The 
radius and surface brightness of each bin are computed as weighted averages of 
the respective input values, where the weights are set to the errors in the 
original light profiles.  Conversely, the enclosed brightness of each bin is 
simply set to the value for the outermost isophote contained therein.  The 
differential and integrated $S/N$ for each bin are used, respectively, to 
define the statistical uncertainty in its surface and enclosed brightness.  
Lastly, the size of the central-most bin for each galaxy is set to match the 
maximum seeing disk amongst all bands being modelled.  This ensures that the 
colours in these regions are accurately measured.  The seeing values for the 
SHIVir data were taken from \cite{M11}.  Note that the binning algorithm 
employed here is an adaptation of the one described in \cite{R11a}.

For each SHIVir galaxy, the above preparations yield refined luminosity 
profiles and non-parametric curves of growth for any desired combination of the 
$grizH$ filters.  From the former, we measure the necessary colours for each 
radial bin and compare these data against a set of MLCRs to obtain a profile of 
\MLiA.  Combining this profile with its $i-$band curve of growth gives us the 
cumulative \MStrB profile and thus the 1D mass within the $R_{23.5,i}$ 
aperture.  At the same time, the integrated colours of the galaxy are extracted 
from the curves of growth for the same aperture.  These colours are then 
modelled with the same set of MLCRs to arrive at the 0D mass.  The major 
distinction between our 0D and 1D masses is that the latter account for the 
luminosity and colour gradients in our galaxies, whereas the former simply 
represent their luminosity-weighted sum.

\subsubsection{2D masses}\label{sec:mod-2d}

The starting point of our 2D approach to \MStrB measurements is the $grizH$ 
imaging for a galaxy.  Recall that, while SHIVir's optical imaging comes 
entirely from the SDSS, the NIR imaging originate from a number of sources.  
For the latter, we use preferentially the ULBCam images for several reasons: 
(i) the field of view ($\sim8.5^{\prime}\times8.5^{\prime}$) permits \SigMB to 
be measured to large radii, with good sky sampling; (ii) the photometric depth 
is high ($\mu_{\text{H}} <$ 24 mag arcsec$^{-2}$); (iii) the sensitivity and 
spatial distortions have been mapped to high precision; (iv) the sub-sample 
spans all major galaxy types; (v) the photometric calibration is homogeneous 
and robust; and, (vi) the small pixel scale (0.25\arcsec) ensures high spatial 
resolution for our mass maps.  The major disadvantage of the ULBCam imaging is 
that it lacks an astrometric calibration, which could compromise the accuracy 
of the measured pixel-by-pixel colours.  To overcome this, we computed 
astrometric solutions\footnote{http://nova.astrometry.net/} for our optical and 
NIR images based on the positions of foreground stars.  We have used our 
optical images in this operation for improved homogeneity.

Other steps are required before a galaxy's colour and stellar mass maps can be 
computed.  The first is to resample the selected images to the largest pixel 
scale therein (if necessary)\footnote{This is done with the SWarp software 
package -- http://www.astromatic.net/software/swarp.  The output from SWarp has 
fixed dimensions for each band and is centered on the galaxy, so that adequate 
sky samples may be later chosen and pixels are properly registered between 
images.}.  Like Z09, we do not correct the images to a common PSF since this 
would corrupt their noise properties and thus invalidate our later use of 
smoothing.  Next, we mask foreground stars and background galaxies that fall 
within the vicinity of the target.  While we generally base our masking upon a 
single image, we make improvements as necessary for light leaked out in other 
bands.  After masking, we place five equal-sized boxes ($N$ = 100 pixels) 
outside the galaxy, in sourceless regions, to measure the background and its 
variation in each image.  The background, which is computed as the mean pixel 
value amongst the boxes, is subsequently subtracted off.  These steps leave us 
with a set of co-registered, background-subtracted images with uniform 
resolution.

As in our 1D masses, we prefer that each resolution element included in our 2D 
masses meet a certain $S/N$ threshold.  To this end, we use the technique of 
adaptive smoothing to boost the $S/N$ of pixels that fall below this 
threshold.  The appeal of this technique is that it preserves the spatial 
sampling of the original images.  However, it comes at the cost of correlating 
the noise between adjacent pixels in the output.  Adaptive spatial binning 
techniques, such as Voronoi tesselation \citep{CC03}, avoid this problem but 
produces images with irregular renditions of the 2D information.  Since these 
irregularities would complicate our modelling of the pixel-to-pixel colours and 
extraction of 2D masses (described below), we prefer to raise the $S/N$ of 
pixels in our images, where necessary, via smoothing.

We smooth our images with the {\tt ADAPTSMOOTH} code \citep{Z09a}, assuming 
background-dominated noise.  While this assumption ignores the Poisson 
contribution from the source, the pixels that require smoothing are generally 
located in regions of low surface brightness, where the background is still 
dominant (S. Zibetti, {\it priv. comm.}).  Moreover, determining the source's 
contribution to the flux in each pixel requires knowledge of the detector gain, 
which is often poorly constrained for stacked images, as in our SHIVir NIR 
data.  We set the noise for each image as the rms variation in the background 
level and threshold $S/N$ of 20, as per the resolution of our MLCRs.  {\tt 
ADAPTSMOOTH} is initially run on selected images to obtain a set of smoothing 
masks that indicate the size of aperture each pixel requires in order to meet 
the desired $S/N$ threshold in all bands.  These masks are then maximally 
stacked to generate a master mask.  We re-run {\tt ADAPTSMOOTH} on the original 
images with this master mask in order to apply a consistent smoothing solution 
to every pixel across all filters.  Since smoothing corrupts the noise 
properties of the output images, we conservatively assume that the $S/N$ in 
each pixel is the threshold value, thus maximizing the errors.  For all {\tt 
ADAPTSMOOTH} runs, we limit the smoothing aperture to a radius $\le$20 pixels.  
Any pixel requiring a larger aperture in any band is clipped within all output 
images.

After smoothing, the images are photometrically calibrated at each pixel.  
Calibration zeropoints to the AB magnitude system were kindly provided by M. 
McDonald.  We also correct the images for Galactic extinction, according to 
\cite{SF11}.  These final preparations leave us with a set of masked, 
background-subtracted, and calibrated galaxy images which have a common spatial 
sampling, registration, and minimum $S/N$ per pixel.

With calibrated surface brightness maps in hand for a select set of filters, we 
construct the stellar mass map of a galaxy by first computing the minimum 
number of colours for each pixel (\Se{mod-m2l}) and transforming those data 
into a map of \MLiB through a set of MLCRs.  The mass map follows by simply 
multiplying the $i-$band and \MLiB maps together.  A 2D \MStrB is then 
extracted from the mass map by a curve-of-growth analysis.  Specifically, we 
integrate the mass map using a fixed set of elliptical apertures (performed 
using the {\tt XVISTA} software package; see \citealt{C96}; \citealt{M11}; 
\citealt{H12}) that match the ellipticity and position angle of the aperture 
from which the 0D/1D masses were measured.  This ensures that extraction 
methods do not introduce any observed differences between the three versions of 
our \MStrB measurements.


\section{Results \& Discussion}\label{sec:r&d}

We have described in the previous section our method for modelling the colours 
of galaxies in terms of their underlying mass-to-light ratio.  We can now use 
this method to quantify the systematics in \MStrB measurements obtained via 
MLCRs.  While previous studies have addressed various aspects of this topic 
from a theoretical (idealised) perspective, we use a representative sample of 
observed local galaxies to focus on the following sources of scatter: (i) SED 
fitting versus MLCRs; (ii) choice of priors; (iii) choice of constraining data; 
and (iv) number of spatial dimensions considered.  Testing MLCRs with real 
galaxies should reveal systematics, if any, that are not typically accounted 
for by the assumed priors.  Note that our discussion below is largely based on 
the BC03 MLCRs.

\subsection{The Correspondence Between MLCRs and SED Fitting}\label{sec:r&d-mlcr}

In \Se{mod-mock} we used mock galaxies to assess the accuracy of \MStrB 
measurements from SED fitting under various conditions of data sampling and 
quality.  Assuming perfect model priors, we found that the scatter in recovered 
\MStrB is approximately constant, with $\sim$0.1 dex, regardless of the filter 
combination within the UV-NIR range and whether $S/N$ = 10 or 20 is achieved 
(see \Fig{sed-vs-true})\footnote{In particular, \MStrB measured from SEDs 
sampled with only the $gi$ filters fare equally well as those determined using 
filters throughout the UV-NIR wavelength range.}.  Since SED-based \MLB 
measurements should be less biased than those from MLCRs (T11), by design, this 
result provides us with a useful benchmark for evaluating the quality of 
MLCR-based results, which is the aim of this section.

We take two complementary approaches to test the correspondence between SED 
fitting and MLCRs, both of which focus on the recovery of \MStrA.  The first 
approach uses mock galaxies.  The advantages of this approach are that: (i) the 
mass to be recovered is known in each case, and (ii) randomly-drawn mock 
galaxies should populate a significant fraction of the observational space 
spanned by a given MLCR set.  Therefore, mock galaxies offer a most direct way 
to uncover any biases introduced by MLCRs.

\begin{figure*}
 \begin{center}
  \includegraphics[width=0.9\textwidth,trim=0 150 0 150]{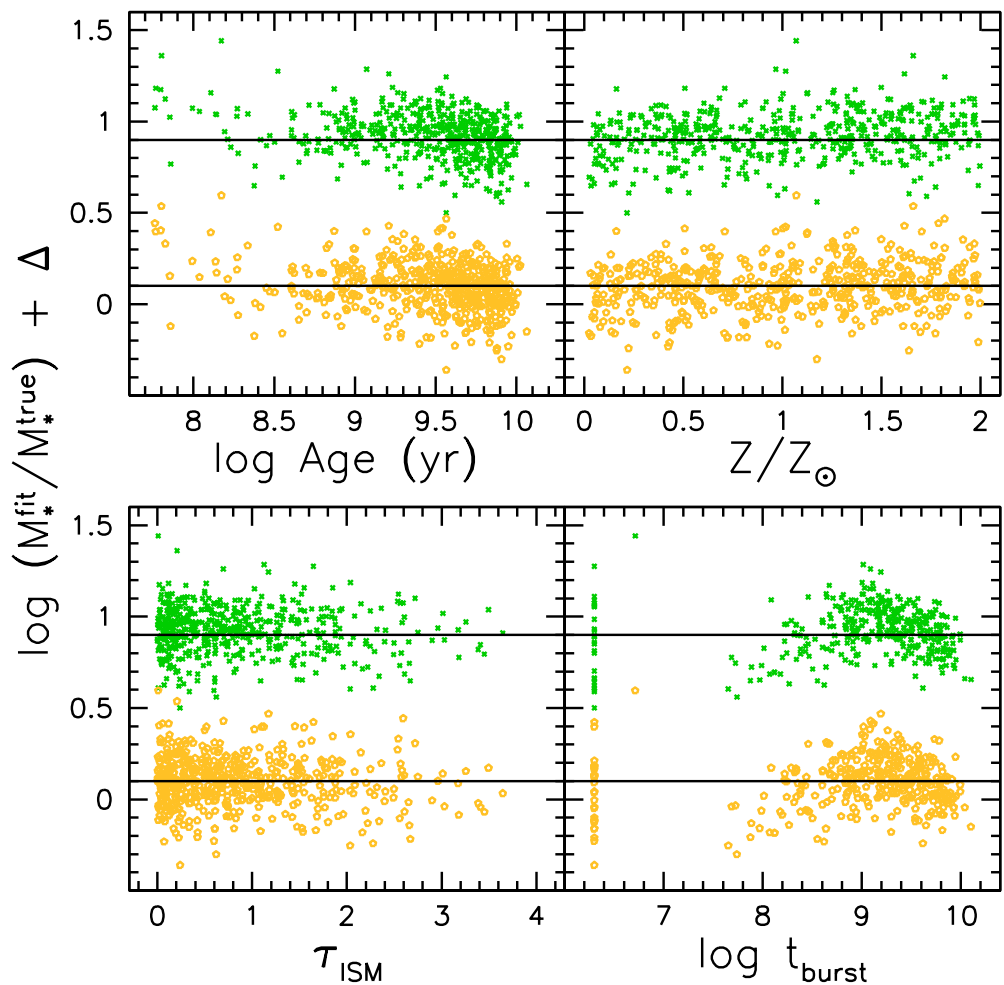}
  \caption{As in \Fig{sed-vs-true} but for \MStrB measured via MLCRs (rather 
  than SED fitting) and SEDs sampled with the $grizH$ (green) and $gi$ (orange) 
  filters, and $S/N$ = 20.  In both cases we find that the residuals have a 
  distribution with zero mean and 0.13-0.14 dex of dispersion.  Similar results 
  are found for the $giH$ and $griH$ filter combinations.}
  \label{fig:mlcr-vs-true}
 \end{center}
\end{figure*}

For our mock galaxy sample, we draw 500 stellar SEDs from the MAGPHYS/BC03 
library.  In \Figure{mlcr-vs-true} we assess the accuracies of the MLCR-based 
masses and trends therein with galaxy parameters under the same scenarios of 
data breadth and quality as assumed for the green and orange points in 
\Fig{sed-vs-true} (\ie $grizH$ and $gi$ band combinations, $S/N$ = 20).  For 
simplicity, we use the same layout and colour scheme as before.

\begin{figure*}
 \begin{center}
  \includegraphics[width=0.9\textwidth]{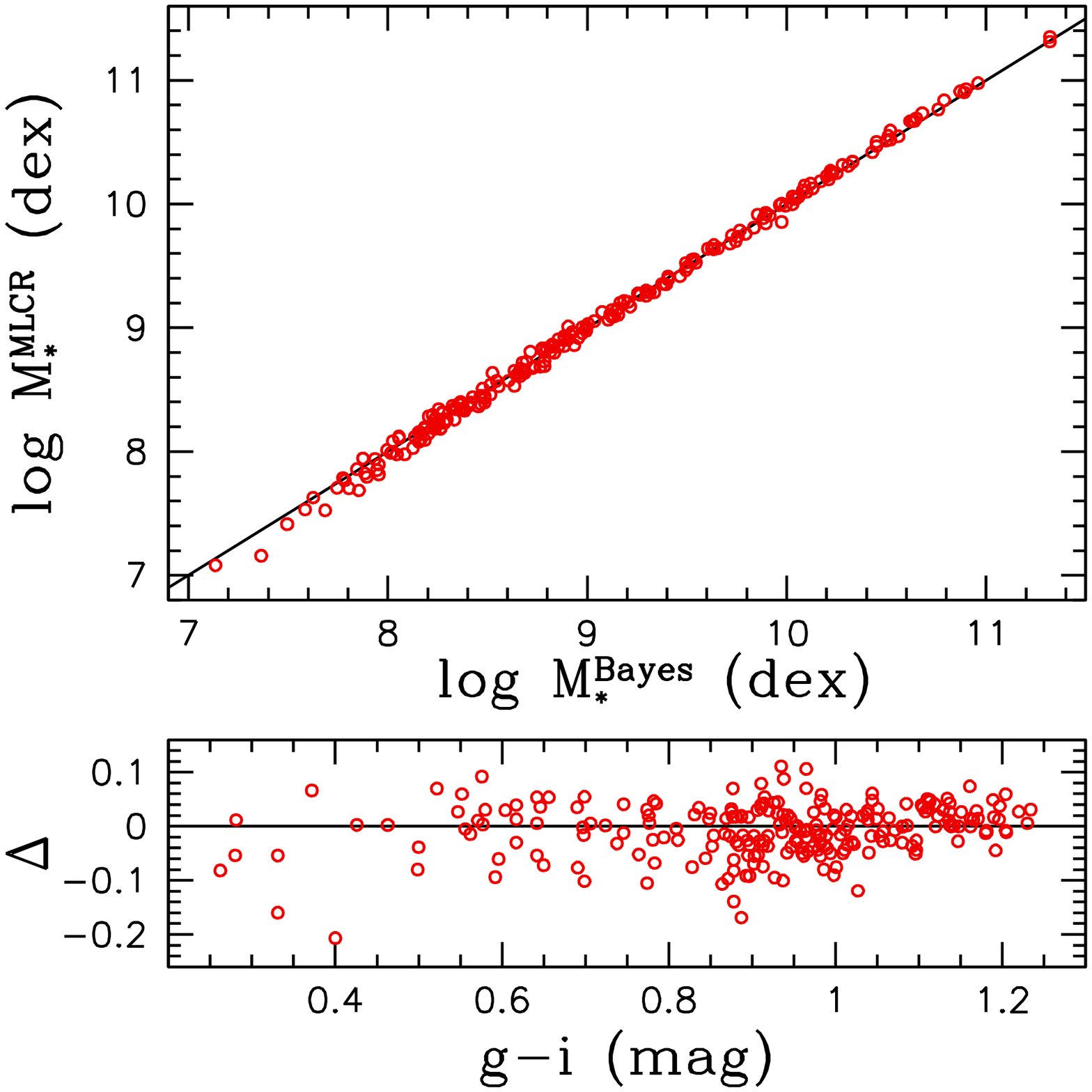}
  \caption{(top) \MStrB estimates from our multi-MLCR method 
  ($M_*^{\text{MLCR}}$) versus those from Bayesian SED modeling 
  ($M_*^{\text{Bayes}}$).  Both sets of masses are derived using the 
  MAGPHYS/BC03 library and the integrated $grizH$ SEDs of SHIVir galaxies.  
  (bottom) Residuals, computed as $\Delta =$ log $M_*^{\text{MLCR}}$ - log 
  $M_*^{\text{Bayes}}$, are plotted as a function of $g-i$ colour.  The 
  correspondence between the two techniques is clearly excellent over the full 
  range of colours probed by our dataset ($\bar{\Delta}$ = -0.01 dex; $\sigma$ 
  = 0.05 dex).}
  \label{fig:mlcr-vs-sed}
 \end{center}
\end{figure*}

\Fig{mlcr-vs-true} shows that our MLCR method yields residual distributions 
that closely resemble those seen in \Fig{sed-vs-true} and, as before, we see 
hints of trends in the residuals as a function of metallicity and \tBurA.  
Moreover, the residuals from our method have nearly identical properties to 
those that were found for SED fitting, with a mean value of (essentially) zero 
and dispersion of 0.13-0.14 dex.  Similar results are obtained when other 
combinations of the SHIVir bands are used (\ie $giH$ and $griH$).  This 
complements the work of GB09, who found that \MLB could be satisfactorily 
recovered for mock galaxies using the $g-i$ colour, except those characterized 
by intermediate-age populations and bursty SFHs (where \MLB was biased to 
higher values by 0.15 dex, in the median).  For galaxies well-described by the 
MAGPHYS priors then, we infer that there is essentially no difference between 
the masses measured via SED fitting or MLCRs.  Furthermore, \Fig{mlcr-vs-true} 
indicates that minimal observational constraints are needed to achieve an 
accuracy of 33\% in the \MStrB measurements for such galaxies.

The correspondence between SED fitting and MLCRs for mock galaxies was to be 
expected since we drew their SEDs from the same SPS model library that was used 
to build our MLCRs.  Our second approach to testing this correspondence 
overcomes this bias by working with real galaxies.  In all likelihood, real 
galaxies populate \MLA-colour space differently than our synthetic models, in 
which case the two methods should yield discrepant results, as demonstrated in 
T11.  A corollary to this is that a sample of real galaxies may not span the 
same range of colours as our models, in which case our second approach can only 
inform us of the correspondence between SED fitting and MLCRs over a limited 
interval.  SHIVir galaxies, for instance, may be biased to redder colours since 
they inhabit a cluster environment.  Still, the outcome of this second approach 
complements the previous one and will provide some context to forthcoming tests.

\Figure{mlcr-vs-sed} compares \MStrB for SHIVir galaxies obtained from MLCRs 
and SED fitting.  Note that this comparison assumes that SED fitting yields the 
most accurate masses for real galaxies.  Both sets of \MStrB were determined 
from the integrated $grizH$ SEDs of these galaxies -- the results for cases 
involving other bandpass combinations are rather similar.  The top panel of 
\Fig{mlcr-vs-sed} indicates that the two methods are in excellent agreement, 
with a mean residual of -0.01 dex and 0.05 dex of dispersion, while the lower 
panel shows that the residuals are independent of galaxy colour, although the 
sampling for $g-i <$ 0.6 is admittedly poor.  This is at odds with T11, who 
emphasised the bias imparted by MLCRs on \MStrA, but our use of multiple 
colours explains this difference.  For instance, if we used the $g-i$ colour 
only, we would tend to overestimate \MLiA, in agreement with T11 (see 
\Fig{x-vs-grizH}).

The above tests demonstrate the two following points about our multi-MLCR 
method: (i) like SED fitting, it recovers the known \MStrB of mock galaxies to 
better than $\pm$0.15 dex; and (ii) it reproduces \MStrB from SED fitting to 
real galaxies to within $\pm$0.05 dex.  The increased accuracy in the latter 
case is likely caused by the smaller colour range spanned by the SHIVir 
galaxies, compared to our mock dataset.  These findings indicate that our 
multi-MLCR method may be counted on to yield fairly accurate results.

\subsection{The Underlying Priors of MLCRs}\label{sec:r&d-priors}

We have shown in the previous section that MLCRs produce, on average, 
equivalent results to SED fitting.  However, this does not entail that MLCRs 
and how they are used are completely free of systematics.  Since an MLCR set is 
constructed from an SPS model library, the priors underlying the model 
ingredients or how the models are implemented can contribute significantly to 
the error budget.  Specifically, any SPS model library parameter that affects 
the relationships between colour, age, metallicity, reddening, and \MLB will 
govern our results.  For instance, omitting a burst component from model SFHs 
will bias \MLB to higher values.  Given the well-studied role that model priors 
play in SED-based \MStrB \citep{C13,C14}, we now examine how these effects are 
manifested in MLCR-based \MStrA.  We omit the choice of IMF from the following 
since its effect, while significant ($\Delta$ \MStrB $\sim$0.3 dex), is well 
understood in most cases \citep[\eg][]{vD08,LS09,C14}.

\begin{figure*}
 \begin{center}
  \includegraphics[width=0.9\textwidth]{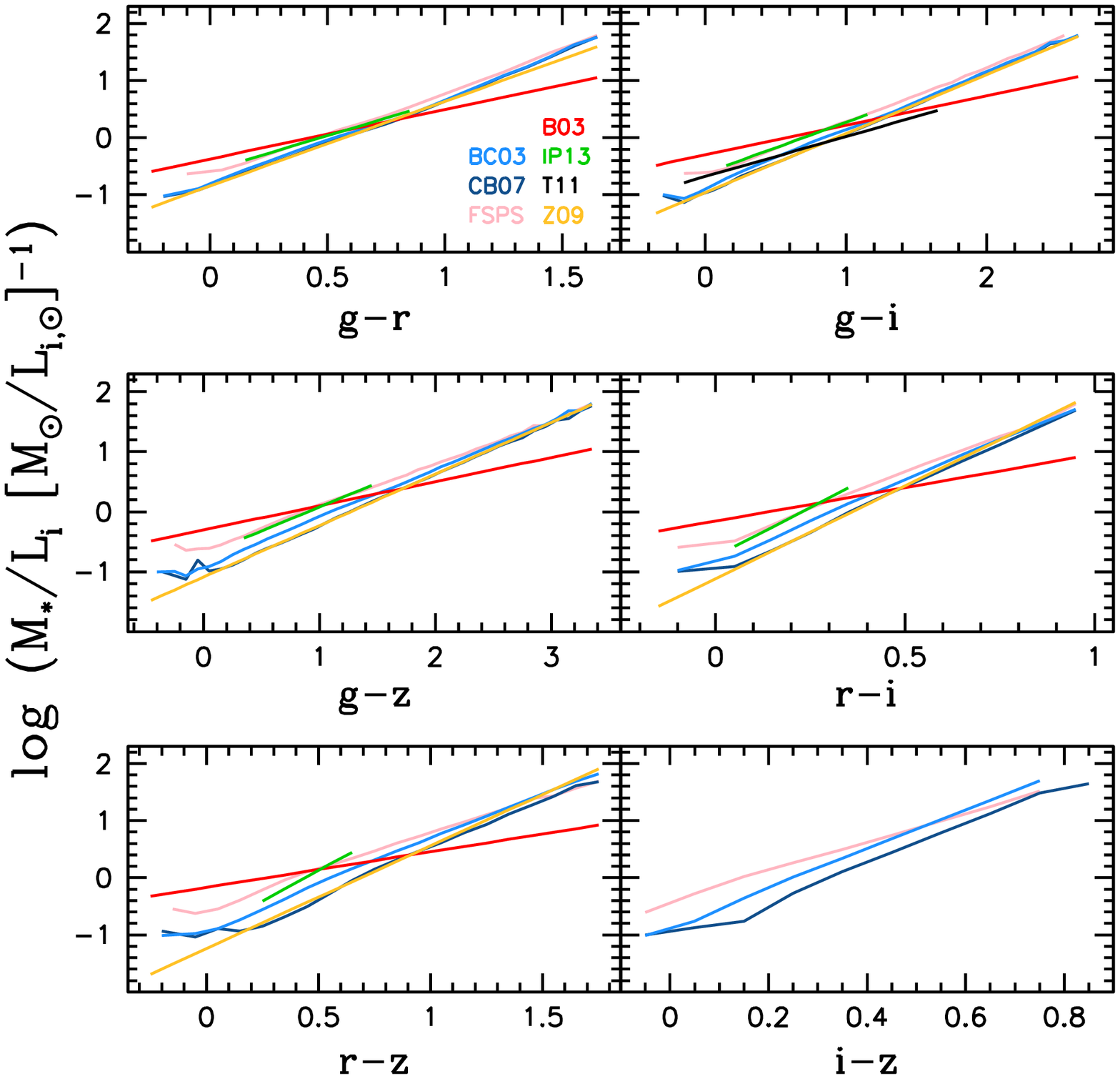}
  \caption{As in \Fig{mlcr-demo} but now showing the MLCRs predicted for SHIVir 
optical colours from libraries of different SPS models (BC03; CB07; 
\citealt{C09}, hereafter FSPS) or related work in the literature 
(\citealt{B03}, \citealt{Z09b}, \citealt{T11}, \citealt{IP13}; hereafter B03, 
Z09, T11, and IP13, respectively).  Note that the B03 and T11 MLCRs are 
constrained empirically by SED fits to real galaxies while all others are fully 
theoretical -- the BC03, CB07, and FSPS relations are of our own making.  Also, 
the B03 relations have been reduced by 0.15 dex to change the underlying 
initial mass function (IMF) to the Kroupa/Chabrier form used by all other MLCRs 
shown here and the IP13 relations correspond to their exponential star 
formation history model.}
  \label{fig:comp-mlcr-opt}
 \end{center}
\end{figure*}

\begin{figure*}
 \begin{center}
  \includegraphics[width=0.9\textwidth]{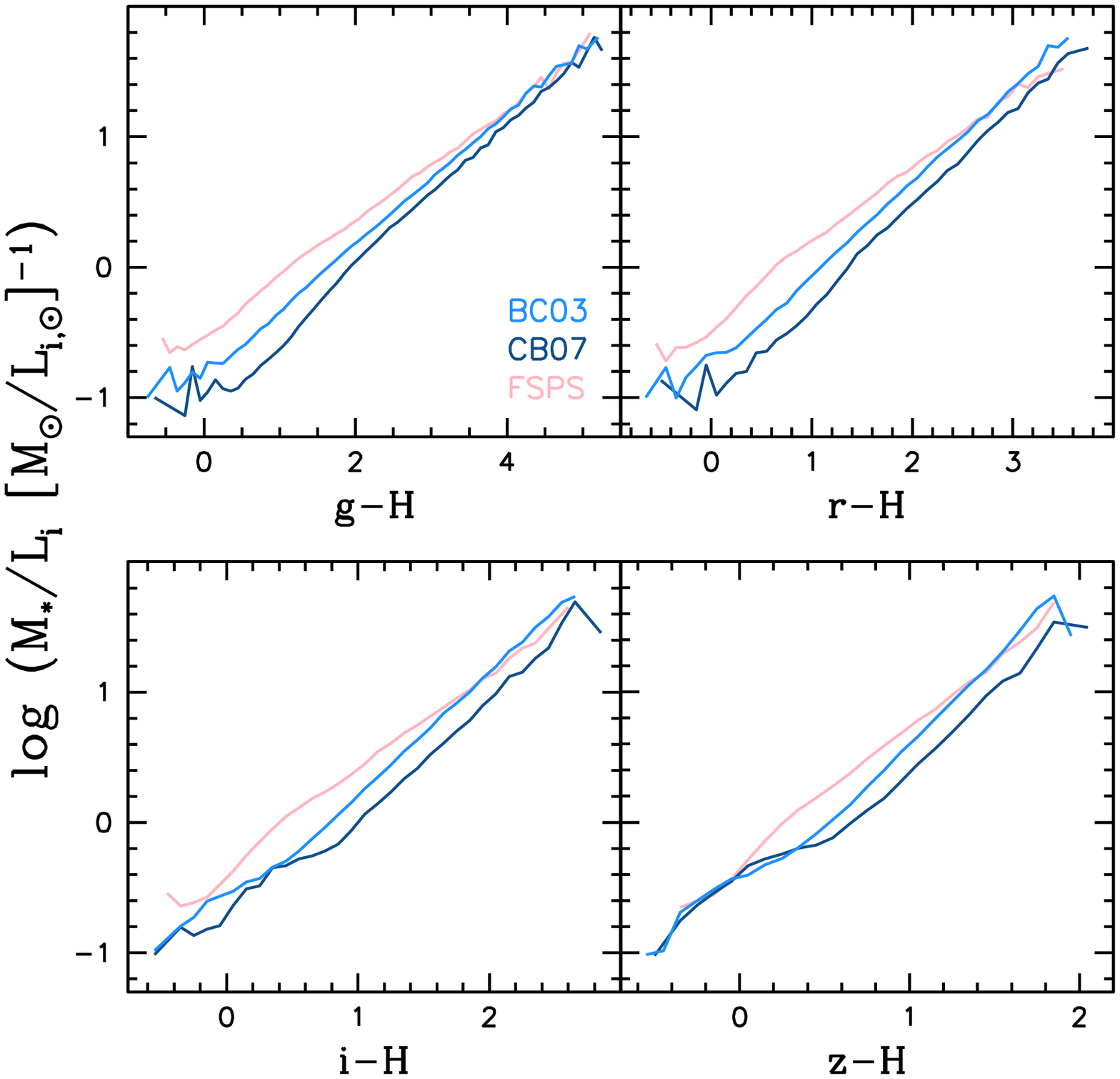}
  \caption{As in \Fig{comp-mlcr-opt} but now comparing MLCRs predicted for 
SHIVir optical-NIR colours, where available.}
  \label{fig:comp-mlcr-opt+nir}
 \end{center}
\end{figure*}

A straightforward approach to assess the effect that priors have on MLCR-based 
\MStrB is to compare MLCRs calibrated with different model libraries.  This is 
achieved in \Figures{comp-mlcr-opt}{comp-mlcr-opt+nir}, where the MLCRs are 
either of our own creation or drawn from the literature.  These plots are 
structured similarly to \Fig{mlcr-demo} though the first focusses on optical 
colours accessible through SHIVir and the second on optical-NIR colours.  Our 
own MLCRs correspond to calibrations based on the BC03 and CB07 versions of the 
MAGPHYS library, as well as the SPS model of \citet[hereafter FSPS]{C09}.  FSPS 
provides the proper machinery to create libraries like MAGPHYS.  With it, we 
have computed 50,000 SEDs which adhere as closely possible to the MAGPHYS 
priors.  Adhering to these priors allows us to gauge the impact that the choice 
of SPS model has on MLCRs.  Two exceptions to our adherence include the shape 
of the effective absorption curve for birth cloud dust and the treatment of 
bursts of star formation.  In MAGPHYS, the optical depths of ambient and birth 
cloud dust are parameterized as power laws in wavelength to different exponents 
(-0.7 and -1.3, respectively), while bursts occur randomly throughout each 
model's lifetime and can last anywhere from 0.03-0.3 Gyr.  In FSPS, both dust 
components are simultaneously described by a single power-law relation, albeit 
with an adjustable exponent, while only one instantaneous burst can be added to 
any model's SFH.  For those readers interested in incorporating our BC03 and 
FSPS MLCRs into their own work, we present tabulations of linear least-
squares fits to these relations in the Appendix.

For the literature MLCRs in \Fig{comp-mlcr-opt}, we show the calibrations of 
B03, Z09, T11, and IP13.  We omit the MLCR sets of BdJ01 and MS14 given their 
Johnson-Cousins band calibrations, whereas our focus is on the SDSS + 2MASS 
suite.  GB09 did not tabulate the MLCRs used in their work but we suspect that 
they are well-matched to those of Z09.  The number of literature MLCRs shown in 
\Fig{comp-mlcr-opt} depends on the colours examined here.  For instance, the 
T11 MLCR was only defined in terms of $g-i$.  Whenever possible, we have 
restricted the range of colours spanned by the literature MLCRs to those 
advocated by their respective authors.  This is why, for example, the T11 and 
IP13 MLCRs do not span the full baselines of the panels in which they appear.  
An important distinction about the trends shown in \Fig{comp-mlcr-opt} is that 
those from B03 and T11 were constrained by their authors using SED fits to real 
galaxies, while all others are purely theoretical.  Lastly, the B03 relations 
have been reduced by 0.15 dex to correct their choice of a ``diet-Salpeter IMF" 
to a \cite{C03} IMF.

In \Fig{mlcr-demo} we demonstrated that, in the case of our BC03 MLCRs, \MLiB 
follows a largely monotonic relationship with each one of the colours defined 
by the $grizH$ bands.  \Figs{comp-mlcr-opt}{comp-mlcr-opt+nir} show that this 
behaviour is likewise reflected, albeit to different degrees, by other sets of 
MLCRs.  The reason for this is simple: as the age, metallicity, and/or 
reddening of a stellar population increases, so do its \MLB and colours.

\Fig{comp-mlcr-opt} highlights that the best agreement between the optical 
MLCRs occurs with $g-r$ and $g-i$, while all other colours are characterized by 
notable discrepeancies.  Not surprisingly, the CB07 and Z09 MLCRs compare very 
well in all panels since they were both derived from the MAGPHYS/CB07 library.  
B03 notwithstanding, these two calibrations also predict the lowest \MLiB 
values for any colour; we therefore focus the following comparisons around the 
CB07 MLCRs.  All other MLCRs, except those of B03 and T11, usually agree better 
with the CB07 calibration towards redder colours (\ie older, more metal-rich, 
and/or highly-reddened populations), but the improvement for IP13 is admittedly 
marginal.  Towards bluer colours (\ie younger, more metal-poor, and/or 
dust-free populations), these same MLCRs can differ from the CB07 calibration 
by as much as $\sim$0.3 dex.  The FSPS MLCR, on the other hand, predicts a 
higher \MLiB by 0.5 dex for the bluest colours.  The absolute discrepancy 
between the T11 and CB07 MLCRs falls within the range $<$0.3 dex at all colours.

The above observations from \Fig{comp-mlcr-opt} indicate that the systematic 
error in MLCR-based \MLB varies as a function of colour.  The cause of this 
colour-dependent error must be due, in part, to the different physics encoded 
by their underlying priors.  It is noteworthy that the B03 and T11 MLCRs 
possess the shallowest slopes, since these are the only relations shown in 
\Fig{comp-mlcr-opt} that were calibrated using real galaxies.  To what extent 
this difference reflects a genuine flaw in theoretical MLCRs is unclear given 
the marked differences in the priors between them and the 
empirically-constrained MLCRs.  For instance, T11 did not include bursts in 
their model SFHs and B03 limited the strength of such events to $\le$10\% by 
mass, whereas the MAGPHYS library incorporates a variety of bursty SFHs.  
Furthermore, B03 omitted reddening from their models, while T11 and MAGPHYS 
invoked the \cite{C00} and \cite{CF00} extinction laws, respectively.  Teasing 
out the relative contributions of priors and the real galaxy distribution in 
the \MLA-colour plane to the slopes of the B03 and T11 MLCRs is non-trivial and 
lies beyond the scope of this work. Furthermore, this comparison between 
empirical and theoretical MLCRs can only inform us on the systematics of 
modelling integrated light.  Investigations like B03 and T11, though based on 
spatially-resolved data, have yet to be performed.

Turning to \Fig{comp-mlcr-opt+nir}, we find that the systematic errors incurred 
by using optical-NIR colours are often much worse than those associated with 
optical colours alone.  This contradicts the popular belief that NIR data yield 
the most accurate \MLB measurements.  We confirm Z09's claim that the 
optical-NIR MLCRs for the MAGPHYS/BC03 library tend to exceed those of the 
MAGPHYS/CB07 library, often by as much as 0.3 dex.  Since the main change to 
the CB07 model was to increase the NIR flux of TP-AGB stars \citep{B07}, we 
attribute this difference to the role of this stage of stellar evolution to the 
error budget.  Supporting this interpretation is the fact that these SPS models 
are implemented identically in the MAGPHYS library, and therefore any 
differences between their MLCRs must be rooted in their most basic ingredients.

\begin{figure*}
 \begin{center}
  \includegraphics[width=0.9\textwidth]{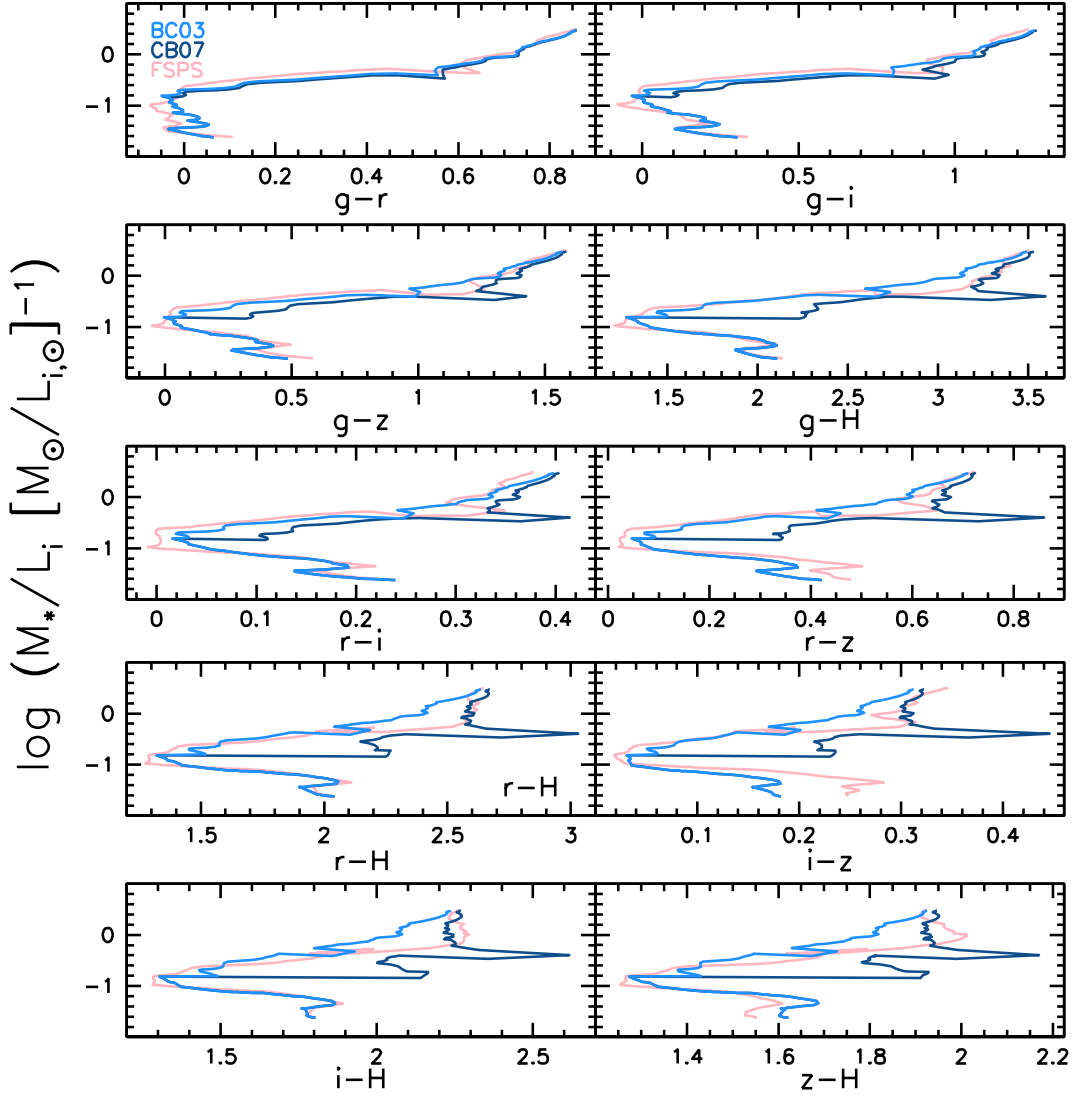}
  \caption{As in \Fig{comp-mlcr-opt} but now comparing the MLCRs based on SSPs 
of solar metallicity drawn from the BC03, CB07, and FSPS SPS models, and for 
all SHIVir colours simultaneously.  Similar trends are also found amongst these 
SSPs at lower metallicities (\eg $Z$ = 0.004).}
  \label{fig:comp-mlcr-ssp}
 \end{center}
\end{figure*}

Although significant, the differences between the BC03 and CB07 optical-NIR 
MLCRs pale in comparison to those between the CB07 and FSPS MLCRs, which can 
exceed 0.6 dex.  To help identify the source of this discrepancy, we show in 
\Figure{comp-mlcr-ssp} the MLCRs corresponding to the simple stellar 
populations (SSPs) of the BC03, CB07, and FSPS models, at solar metallicity.  
Since the CB07 and FSPS models incorporate the same isochrones 
\citep{MG07,M08}, we expect that the MLCRs for their SSPs should follow each 
other.  Instead, the CB07/SSP MLCRs abruptly diverge from the FSPS/SSP MLCRs to 
redder optical-NIR colours at an age of $\sim$0.1 Gyr, and remain 
well-separated for another $\sim$1.5-1.6 Gyr.  During that time period, the 
FSPS/SSP MLCRs actually follow the BC03/SSP relations.  These behaviours 
qualitatively persist at lower metallicities as well (\eg $Z$ = 0.004).  The 
above age range is well-known as the period in which TP-AGB stars are present 
in SSPs.  Therefore, the TP-AGB treatment must be contributing to the observed 
differences between the CB07 and FSPS optical-NIR MLCRs too.  However, 
\Fig{comp-mlcr-opt+nir} shows that the FSPS MLCRs also lie above the BC03 
MLCRs, despite the similar trends exhibited by their SSPs in 
\Fig{comp-mlcr-ssp}.  We suggest then that the much burstier nature of the 
MAGPHYS SFHs helps drive the differences between the FSPS and BC03/CB07 MLCRs 
as well.  The flat prior on the burst mass fraction in MAGPHYS and the fact 
that half of its SFHs incorporate a recent burst suggest that our BC03/CB07 
MLCRs are biased to stellar populations with younger mass-weighted ages than 
our FSPS MLCRs.  Our suggestion is consistent with the rank order of these 
MLCRs in \Figs{comp-mlcr-opt}{comp-mlcr-opt+nir}, in which the BC03/CB07 MLCRs 
predict a lower \MLiB at bluer colours than do the FSPS MLCRs.  Caution must 
therefore be taken when interpreting \MStrB measurements that are based on 
optical-NIR MLCRs.

\subsection{The Impact of SED Sampling}\label{sec:r&d-sed}

\Figs{comp-mlcr-opt}{comp-mlcr-opt+nir} highlighted the systematic penalty 
incurred with the MLCR method through the choice of an SPS model library.  
Particularly striking was the inflated disagreement between MLCR sets when 
optical-NIR colours are included.  However, it is conceivable that, within the 
context of a single MLCR set, the \MLB measured for a galaxy would also vary as 
a function of the data considered.  One reason could be the fact that not all 
colours are equally sensitive to the parameters that establish \MLA, such that 
the use of multiple colours might enhance one's constraining power.  Another 
could be that the fluxes predicted by SPS models within one or more filters are 
inaccurate, which likewise would affect all colours associated therewith.  
Given these concerns, we address in this section whether or not the degree of 
SED sampling contributes meaningfully to the systematics of MLCRs.

\begin{figure*}
 \begin{center}
  \includegraphics[width=0.9\textwidth]{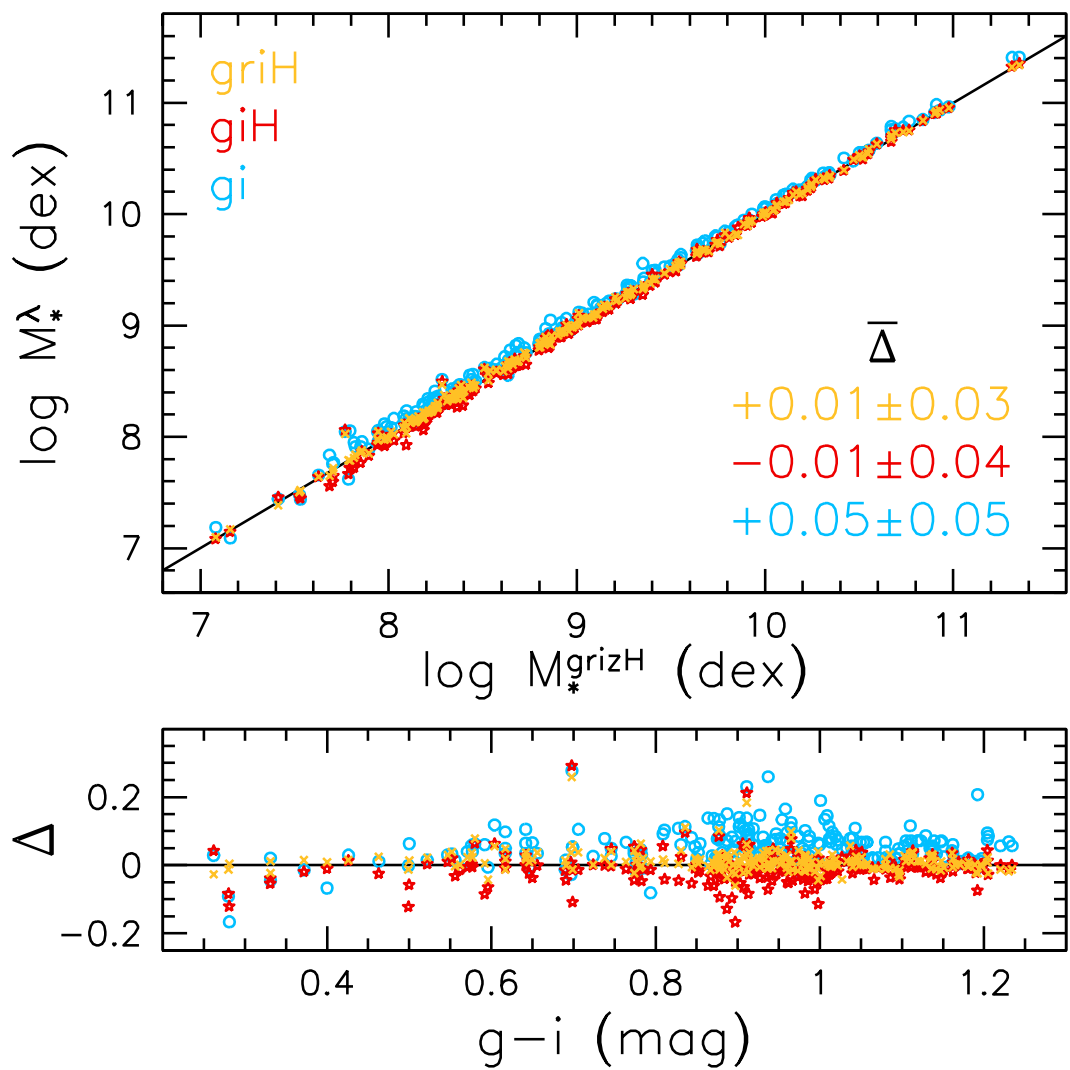}
  \caption{(top) Comparison of \MStrB estimates from our multi-MLCR method for 
  various realizations of the SHIVir dataset.  Masses derived from the filter 
  combinations $gi$, $giH$ and $griH$ are plotted against those measured from 
  the $grizH$ bands.  (bottom) Residuals, computed as $\Delta$ = 
  log$M^{\lambda}_*$ - log$M^{grizH}_*$, are plotted as a function of $g-i$ 
  colour.  The mean and dispersion of each residual distribution are quoted in 
  the top panel.  These values indicate that our method is modestly robust to 
  SED sampling.}
  \label{fig:x-vs-grizH}
 \end{center}
\end{figure*}

To assess the impact of SED sampling, we shall test the robustness of our 
multi-MLCR method under a variety of realizations of the SHIVir data.  We 
compare in \Figure{x-vs-grizH} sets of \MStrB measured for these galaxies via 
our BC03 MLCRs and SEDs sampled with the following filter combinations: $gi$, 
$giH$, $griH$, and $grizH$.  The layout of this figure is identical to 
\Fig{mlcr-vs-sed}) and we use the same combinations of colours as before in the 
fits for each SED realization.  The $gi-$, $giH-$, and $griH-$based \MStrB are 
plotted in the top panel as a function of the $grizH-$based \MStrA, which we 
take to be the most accurate (see \Fig{mlcr-vs-sed}).  In the bottom panel we 
plot the residuals between the former and latter as a function of the $g-i$ 
colour.

\Fig{x-vs-grizH} demonstrates that the $grizH-$based masses of SHIVir galaxies 
are reproduced well by those using the $griH$ and $giH$ band combinations 
instead.  The $gi-$based values, on the other hand, are biased to slightly 
higher values, by 12\% on average, implying that the $g-i$ colour 
systematically overestimates \MLiA, in agreement with the findings of T11.  
In the bottom panel, we see that the residuals from the various realizations of 
the SHIVir dataset are robust to galaxy colour but, as in \Fig{mlcr-vs-sed}, 
this conclusion is tentative for systems with the bluest colours ($g-i <$ 0.6).

The robustness of \MStrB for SHIVir galaxies is explained by how we combine 
\MLiB values when multiple colours are fitted.  Recall that we construct a 
master \MLiB for each object by computing a weighted mean of the values that 
were assigned to its colours.  Our method then naturally assigns greater weight 
to those colours which have less scatter in their MLCRs.  In the case of 
\Fig{x-vs-grizH}, the colour that maintains the robustness amongst the various 
sets of \MStrB is $g-i$.  Indeed, \Fig{mlcr-demo} demonstrates (via the dashed 
lines) how the scatter in \MLiB is minimized for the $g-i$ MLCR, relative to 
the $g-r$, $g-z$, and $g-H$ MLCRs.

Since most sets of \MStrB presented in \Fig{x-vs-grizH} are mutually 
consistent, one may be tempted to conclude that the $r$ and $z$ bands do not 
increase the constraining power of our method.  While it is true that including 
these bands does not change \MLiB significantly, a positive statistical impact 
is made on the assigned errors.  The mean error on \MLiB with the $gi$, $giH$, 
$griH$, and $grizH$ SEDs of SHIVir galaxies, is 0.29, 0.22, 0.18, and 0.15 dex, 
respectively.  Thus while the $g-i$ colour may be counted on to reasonably 
determine \MStrB when other photometry is scarce and/or unreliable, the 
desirable combination of high accuracy and precision is achieved by applying 
MLCRs to well-sampled SEDs.

\subsection{Must Mass Estimates be Spatially-Resolved?}\label{sec:r&d-dim}

We now examine the claim that \MStrB is more accurately measured on the basis 
of spatially-resolved data than integrated fluxes (Z09).  According to this 
claim, the method by which one measures a galaxy \MStrB is most relevant if a 
significant amount of it (\eg $>$10\%) is hidden by bright stars and/or dust.  
Higher-dimensional approaches to measuring \MStrB presumably overcome this bias 
by assigning each galaxy resolution element a \MLB tailored to its colours.  In 
other words, allowing \MLB to vary spatially within a galaxy may be essential 
to ensure an accurate determination of its \MStrA.

The need for a 2D approach to measuring \MStrB was first demonstrated by Z09.  
Using optical and NIR imaging, they found that the \MZerB of spiral galaxies 
could be lower than their \MTwoB by as much as 0.15 dex.  Conversely, the 2D 
approach produced hardly any change in \MStrB for the lone early-type galaxy in 
their sample.  The authors thus claimed that the discrepancy between \MZerB and 
\MTwoB increases towards later galaxy types.  The Z09 sample, however, was 
small and biased to late-type spirals.  We have improved upon this limitation 
by expanding the work of Z09 to a larger, more representative set of local 
(SHIVir) galaxies (see also SS15).  Similarly, the use of a 1D approach (via 
luminosity profiles) may also yield improved \MStrB measurements, by accounting 
for the colour gradients which many galaxies are known to possess \cite{R11a}.  
A 1D approach is also more straightforward and less computationally-expensive 
than its 2D cousin.  We therefore explore the relation between \MZerB and 
\MOneB in this section as well.

\begin{figure*}
 \begin{center}
  \includegraphics[width=0.9\textwidth]{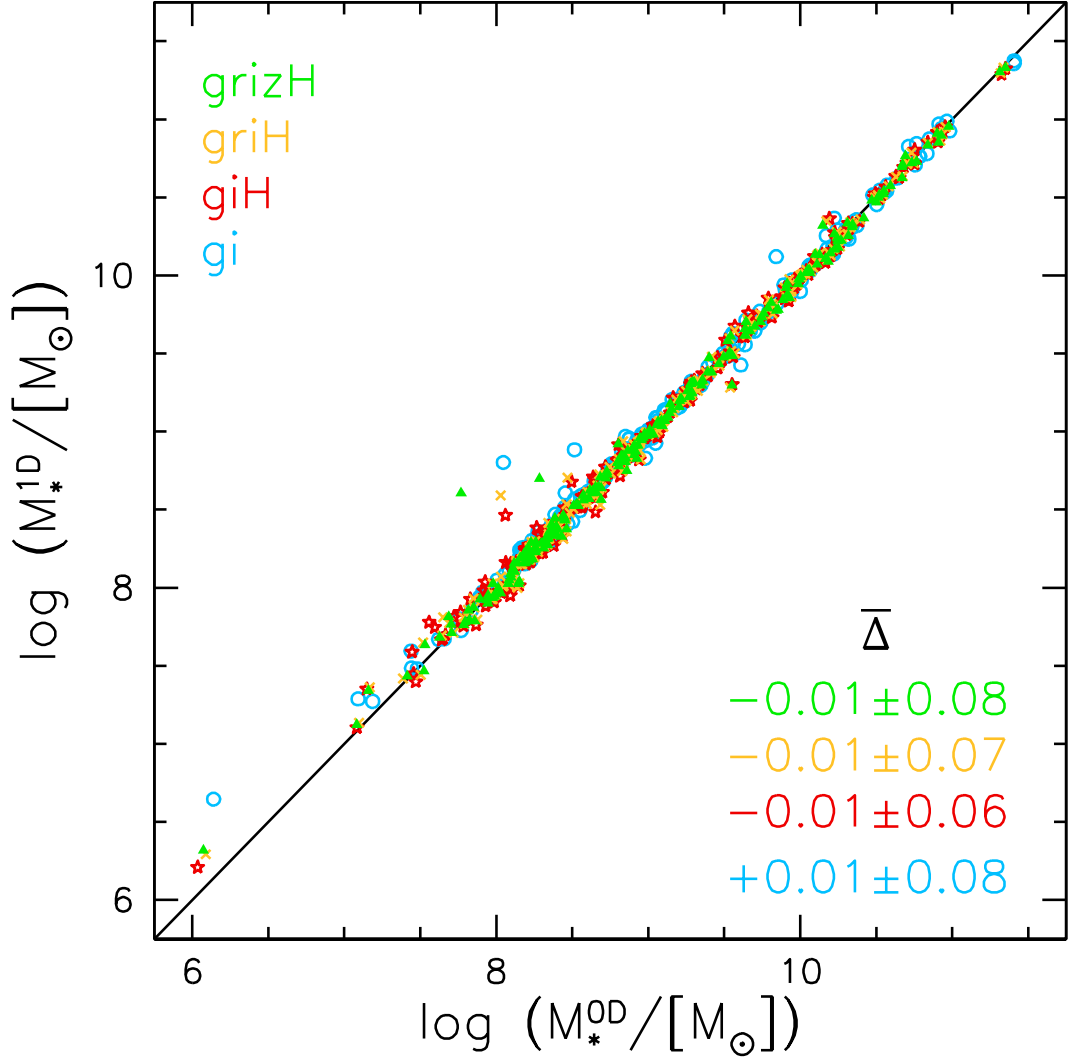}
  \caption{\MStrB estimates from the integration of stellar mass density 
  profiles (\MOneA) versus those from integrated photometry (\MZerA).  Both 
  sets of masses have been derived for various realizations of the SHIVir 
  dataset ($gi$; $giH$; $griH$; $grizH$).  The mean residuals and dispersions 
  (tabulated at lower-right) suggest that, for our sample, colour gradients do 
  not introduce a persistent, significant bias in \MZerA.}
  \label{fig:1d-vs-0d}
 \end{center}
\end{figure*}

\Figure{1d-vs-0d} compares the 0D and 1D \MStrB for SHIVir galaxies based on 
SEDs sampled with either the $gi$, $giH$, $griH$, or $grizH$ bands.  The 
correspondence between the 0D and 1D values is clearly excellent in most cases, 
in the sense that there is no net bias, as evidenced by the residual 
statistics.  The major outliers are understood as being due to cases of extreme 
colour gradients, where a positive gradient would lead to \MOneB $>$ \MZerA and 
vice versa.  Omitting the outliers, we infer that radial variations of \MLiB 
within galaxies, on average, bias 0D measurements by no more than $\pm$20\%.  
Several explanations could apply here: (i) most galaxies have mild colour 
gradients; (ii) as in (i) but only for those colours which possess steep MLCRs 
(\eg $g-r$), or (iii) \MLiB variations occur in regions of galaxies which do 
not enclose large fractions of their total light.  We propose that the third 
explanation applies most often, as earlier work by our group \citep{R11a} 
demonstrated that significant colour gradients are commonly found within the 
SHIVir sample.

\Fig{1d-vs-0d} suggests that colour gradients do not affect \MStrB measurements 
for our galaxies to a large degree.  In addition to the radial coordinate 
though, galaxy colours can vary with azimuth and asymmetrically (\ie 
pixel-by-pixel).  1D masses are exempt from these other possible dependencies 
by virtue of the isophotal fitting that is used to measure luminosity profiles 
of galaxies.  2D masses, however, are sensitive to all three.  To address the 
significance of the latter two dependencies, and the veracity of Z09's 
conclusions, we have derived 2D masses for 18 nearly face-on SHIVir galaxies, 
which is twice the size of Z09's sample\footnote{Computational constraints 
limited us to this sub-sample.  We will present the mass maps for all the 
$\sim$280 SHIVir galaxies in a forthcoming publication.  SS15 have also 
recently computed 2D masses for 67 nearby galaxies.}  Some basic properties of 
our 2D sub-sample are listed in \Table{2d-samp}.  The morphological 
distribution (column 5) show that this sub-sample represents the present-day 
galaxy population quite well, spanning the giant early-types, spirals, and both 
flavours of dwarves (gas-rich and -poor). Their 0D \MStrB should thus encompass 
the full range of biases mentioned above, if indeed present.  

\begin{deluxetable}{rccrlcccc}
 \tabletypesize{\footnotesize}
 \tablewidth{0pc}  
 \tablecaption{Bulk properties of Virgo cluster galaxy sub-sample used for 2D mass map experiment\tablenotemark{1}.}
 \tablehead{
  \colhead{NGC} &
  \colhead{VCC} &
  \colhead{$\alpha_{\text{2000}}$} &
  \colhead{$\delta_{\text{2000}}$} &
  \colhead{Type\tablenotemark{2}} &
  \colhead{$M$\tablenotemark{3}} &
  \colhead{$C_{28}$} &
  \colhead{$\mu_{e}$} &
  \colhead{$R_{e}$\tablenotemark{3}} \\
  \colhead{ID} &
  \colhead{ID} &
  \colhead{(deg)} &
  \colhead{(deg)} &
  \colhead{} &
  \colhead{(mag)} &
  \colhead{} &
  \colhead{(mag arcsec$^{-2}$)} &
  \colhead{(kpc)} \\
  \colhead{(1)} &
  \colhead{(2)} &
  \colhead{(3)} &
  \colhead{(4)} &
  \colhead{(5)} &
  \colhead{(6)} &
  \colhead{(7)} &
  \colhead{(8)} &
  \colhead{(9)}
 }
 \startdata
  $-$ & 0389 & 185.0138 & 14.9614 & dS0 & -18.32 & 3.690 & 22.16 & 1.454 \cr
  $-$ & 0751 & 186.2012 & 18.1950 & dS0 & -17.68 & 3.374 & 22.00 & 0.994 \cr
  $-$ & 0788 & 186.3200 & 11.6053 & dE  & -15.98 & 2.992 & 23.26 & 0.970 \cr
  $-$ & 0816 & 186.4000 & 15.8478 & dE  & -17.70 & 2.665 & 24.46 & 3.210 \cr
  $-$ & 1437 & 188.1396 &  9.1736 & E   & -16.96 & 4.055 & 21.00 & 0.437 \cr
  $-$ & 1440 & 188.1392 & 15.4153 & E   & -17.87 & 5.797 & 22.12 & 0.914 \cr
 4503 & 1412 & 188.0258 & 11.1764 & S0  & -20.76 & 4.701 & 20.83 & 2.308 \cr
  $-$ & 1358 & 187.8467 & 17.2064 & Sa  & -16.91 & 2.679 & 23.23 & 1.361 \cr
 4584 & 1757 & 189.5746 & 13.1100 & Sa  & -18.65 & 3.464 & 22.05 & 1.744 \cr
 4407 & 0912 & 186.6342 & 12.6111 & Sab & -19.64 & 3.411 & 22.07 & 2.919 \cr
 4394 & 0857 & 186.4812 & 18.2142 & Sb  & -21.03 & 4.239 & 22.01 & 4.284 \cr
  $-$ & 2023 & 191.3833 & 13.3325 & Sc  & -17.80 & 1.943 & 23.33 & 2.377 \cr
  $-$ & 0510 & 185.4738 & 15.6458 & Sd  & -17.42 & 3.391 & 23.55 & 1.916 \cr
 4519 & 1508 & 188.3758 &  8.6547 & Sd  & -19.50 & 2.897 & 22.12 & 2.775 \cr
  $-$ & 0583 & 185.6875 & 15.5022 & Im  & -16.83 & 2.384 & 24.04 & 1.758 \cr
  $-$ & 0980 & 186.7967 & 15.8967 & Im  & -17.52 & 2.488 & 23.60 & 2.373 \cr
  $-$ & 1725 & 189.4229 &  8.5586 & BCD & -17.23 & 2.052 & 23.30 & 1.777 \cr
  $-$ & 1804 & 189.9171 &  9.3989 & BCD & -16.35 & 3.427 & 23.70 & 1.324 \cr
 \enddata
 \label{tbl:2d-samp}
 \tablenotetext{1}{ The parameters $M$, $C_{28}$, $\mu_e$, and $R_e$ correspond to $i-$band values.}
 \tablenotetext{2}{ Drawn from the NASA Extragalactic Database.}
 \tablenotetext{3}{Estimated assuming a common distance to all Virgo cluster galaxies of 16.5 Mpc \citep[\ie $m-M$ = 31.087 mag;][]{M07}.}
\end{deluxetable}

\begin{figure*}
 \begin{center}
  \includegraphics[width=0.9\textwidth]{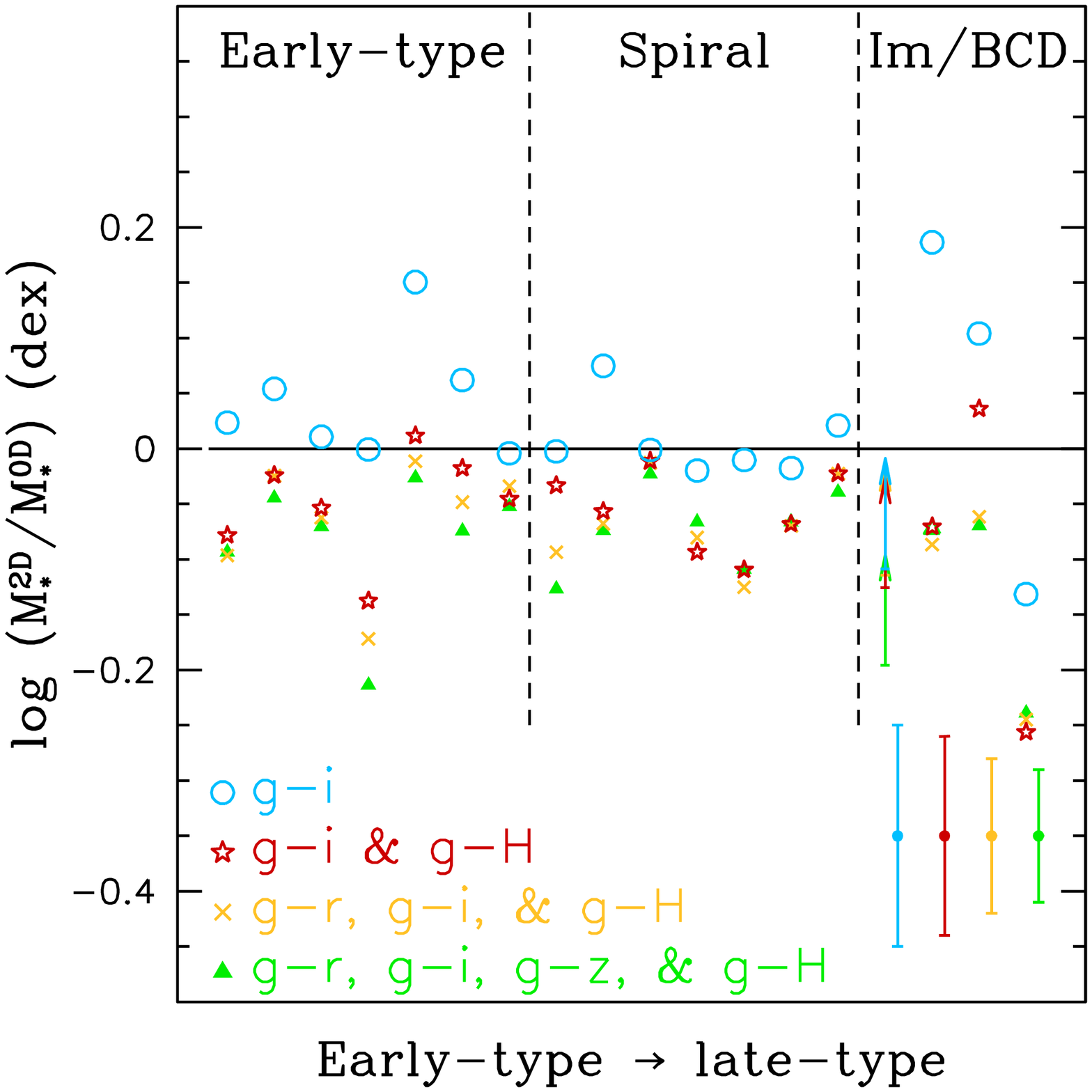}
  \caption{Comparison of \MStrB values derived from integrating the \SigMB maps 
for 18 SHIVir galaxies (\MTwoA) versus \MZerA.  The comparison is plotted as 
the logarithmic difference between these two sets of masses 
[log(\MTwoA/\MZerA)] as a function of morphology and is carried out for SEDs 
sampled with the filter combinations $gi$, $giH$, $griH$, and $grizH$.  
Galaxies are ordered from early-types (dE, dS0, E, and S0) on the left, to 
spirals (Sa$-$Sm) in the middle, and gas-rich dwarves (Im/BCD) on the right, 
with dashed lines separating each of these groups.  The solid line denotes 
\MTwoB = \MZerB and representative error bars are shown in the bottom-right 
corner.}
  \label{fig:2d-vs-0d}
 \end{center}
\end{figure*}

The 0D and 2D galaxy \MStrB are compared in \Figure{2d-vs-0d}.  As for 
\Fig{1d-vs-0d}, we have measured 2D \MStrB (see \Table{2d-samp}) for a variety 
of colour combinations, which amounts to iteratively dropping filters from the 
full SHIVir suite ($grizH$).  Unlike our 0D-1D comparison, we plot the ratio 
log(\MTwoA/\MZerA) instead, where galaxies with \MTwoB $>$ \MZerB lie above the 
horizontal line that marks equality.  Plotting this ratio allows us to order 
galaxies from early-types on the left to late-types on the right, and thus to 
gauge the existence of any correlations with morphology.  Note that each 
vertical cluster of blue, green, orange, and red points represents a single 
galaxy and the dwarf galaxies in our 2D sub-sample appear as the four leftmost 
members of the early-type group and the four members of the "Im/BCD" group.  
Arrows represent cases where only lower limits could be placed on the 2D 
masses, while hollow points denote cases in which 0D masses are unreliable.  
The former occur when {\it ADAPTSMOOTH} cannot build enough $S/N$ out to 
$R_{23.5,i}$ and the latter when one or more colours exceed the ranges of our 
MLCRs.  The error bars shown in the lower-right corner represent the average 
uncertainty in the 0D masses for these galaxies, and should thus be regarded as 
lower limits.

\Fig{2d-vs-0d} shows that our 2D approach produces mixed results with respect 
to 0D values, depending on the data involved.  \MTwoB most often exceeds \MZerB 
when the SEDs of pixels are sampled with the $gi$ band combination.  In the 
median however, this excess is quite small for early-types (dwarves + giants; 
0.02 dex) and zero for spirals, the reason being that many members of these 
groups lie about the \MTwoB = \MZerB line.  While late-type dwarfs possess 
amongst the highest values of log (\MTwoA/\MZerA) when the $gi$ bands are used, 
the small size of their group prevents us from drawing firm conclusions.  
Nevertheless, the offsets seen for some of our Im/BCD's agrees with the fact 
that these galaxies often host pockets of recent star formation, which would 
far outshine their underlying (low surface brightness) stellar 
disks/spheroids.  In other words, we expect the integrated light of these 
galaxies to be biased blue and, accordingly, low \MLA.

Although the offsets between \MZerB and \MTwoB are {\it typically} small for 
most galaxy types when using only the $gi$ bands, they can be most significant 
on a case-by-case basis.  For instance, VCC 1437, an elliptical galaxy with a 
very blue center (most likely due to recent star formation), has \MTwoB $>$ 
\MZerB by 0.15 dex, while three of the Im/BCD galaxies in our 2D sub-sample 
have absolute log (\MTwoA/\MZerA) values of $>$0.1 dex.  Our foremost 
conclusion regarding \MZerB measurements based on optical information alone is 
that internal variations in \MLB may be captured by including an additional 
random component in the error budget for all populations.  Our best estimate of 
the size of this component is 0.05 dex for early-types and 0.03 dex for 
spirals; these values can be refined using a larger and more complete sample in 
future work.

2D masses that include NIR information behave similarly to those based on $gi$ 
bands, except for two important differences: we rarely find positive values of 
log (\MTwoA/\MZerA) and few galaxies cluster about the \MTwoB = \MZerB line.  
The first difference could arise from either an increase in \MZerB or a 
decrease in \MTwoB (or both) as we model more bands than just $g$ and $i$.  
\Fig{x-vs-grizH} shows that, on average, \MZerB measured from the band 
combinations $giH$, $griH$, and $grizH$ are mutually consistent (to within 
$\pm$0.03-5 dex), while $gi-$based masses are $\sim$13\% higher.  This 
behaviour of \MZerB can therefore not explain the observed changes in log 
(\MTwoA/\MZerA); the onus must instead lie with \MTwoA.  Taking ratios of 
\MTwoB between the datasets considered in \Fig{2d-vs-0d} indeed shows that 
\MTwoB typically decreases by $\sim$30\% when the SEDs of pixels are sampled 
with either the $giH$, $griH$, or $grizH$ filters, compared to the case of $gi$ 
SEDs.  Including NIR fluxes in 2D mass measurements thus systematically reduces 
galaxy-wide mean \MLiB values (as in the 0D case; \Fig{x-vs-grizH}) and mass 
maps derived for each of the band combinations $gr$, $gz$, and $gH$ confirm 
that the lattermost yield the smallest values of \MLiB per pixel, on average.  
The discrepancy between optical and optical+NIR masses from our 2D approach 
supports the suggestion of T11 that predictions of current SPS models lack 
consistency across the optical-NIR domain.  Our own results highlight the need 
for further tests of the MLCR method (and SPS models in general), preferably on 
resolved populations, where \MLB may be directly inferred from colour-magnitude 
diagrams.

Unlike with our $gi-$based measurements, the lack of galaxies found in the 
vicinity of the \MTwoB = \MZerB line in \Fig{2d-vs-0d} when NIR data are 
considered implies that 0D values are biased higher than their 2D counterparts 
in such cases.  For early-types and spirals, this bias remains robust amongst 
the $giH$, $griH$, and $grizH$ band combinations, falling between -0.05 and 
-0.07 dex.  The bias for gas-rich dwarfs changes significantly between these 
cases, however, from -0.07, to -0.09, and -0.15 dex, respectively.  In 
addition, log (\MTwoA/\MZerA) measurements scatter about these median offsets 
at roughly the 0.04, 0.05-0.06, and 0.10-0.15 dex level for spiral, early-type, 
and gas-rich dwarf galaxy types.  Again though, these estimates must be 
revisited when larger, homogeneous, and more complete samples of high-quality 
optical + NIR imaging become available for nearby galaxies.  Our experiment 
with 2D masses that include NIR information thus confirms that the use of 
integrated photometry increases the random error in associated \MStrB 
measurements for whole populations, while introducing a non-trivial bias as 
well.  In the worst case, 0D and 2D masses differ by $\sim$0.25 dex when NIR 
data are involved, and $\sim$0.20 dex otherwise.

\Fig{2d-vs-0d} can also be compared to Z09 and SS15, the only prior studies of 
the spatially-resolved approach for measuring \MStrA.  Z09 found that, on a 
galaxy-by-galaxy basis, \MTwoB could exceed \MZerB by as much as $\sim$40-65\% 
(0.15-0.22 dex), depending on the MLCRs and dataset employed; the most 
egregious disagreements occured when they modelled optical-to-NIR SEDs.  
Similarly, SS15 found that, in the most extreme cases, measurements of \MZerB 
missed $\sim$25, 50, and 20\% of the stellar mass in E/S0, spiral, and Irr 
galaxy types.  While we find offsets of similar magnitudes to Z09, they only 
occur in the sense of \MTwoB $>$ \MZerB when we consider optical information 
alone.  With NIR fluxes included, our 2D method typically yields smaller masses 
compared to 0D measurements.  A contributing factor might be that the BC03 
models tend to predict optical-NIR MLCRs of shallower slopes than the CB07 
models (\Fig{comp-mlcr-opt+nir}), thereby leading to milder \MLA gradients.  
This discrepancy raises further awareness of the need for tests of MLCRs and 
SPS models on resolved stellar populations; the biased distribution of points 
in \Fig{2d-vs-0d} pertaining to NIR-based predictions strongly hints at a 
disconnect between SPS models across the optical-NIR domain.  However, it 
should be noted that SS15 measured \MZerB $>$ \MTwoB (by as much as 25\%) for 
nearly one-fifth of their full sample, with these galaxies being distinguished 
by their low-to-medium sSFRs.  Recalling that our sample is drawn from a 
cluster environment and that cluster galaxies tend to exhibit lower sSFRs, 
relative to their field counterparts, the discrepancies we have uncovered may 
be genuine.

Aside from some notable outliers, Z09 also argued in favour of a correlation 
between \MTwoA/\MZerB and galaxy type, whereby later-types have larger ratios.  
They surmised that late-type galaxies, on average, contain more star formation 
and dust than early-types, hence the integrated light of the former should more 
often be biased to lower \MLA.  A closer inspection of their Fig. 13 calls some 
aspects of their claim into question, however.  First, the authors do not 
present any statistical tests for the existence of a correlation (\eg 
Spearman's coefficient) and, second, the correlation disappears when NIR fluxes 
are included in their modelling.  Our own results suggest no such correlation 
exists for any combination of the SHIVir bands.  Instead, our use of a more 
representative sub-sample of galaxies than Z09's has demonstrated that the 
integrated light of all galaxy types, not just late-type spirals, can suffer 
from significant biases.  SS15 also showed (their Fig. 3) that a galxy's sSFR 
appears to be a better diagnotic of these biases.  The fact that we recover a 
dearth of galaxies with \MTwoB $>$ \MZerB may be a consequence of this relation 
and the fact that our sample consists entirely of cluster galaxies.  A 2D 
analysis for a complete sample of galaxies is ultimately needed to establish 
the dependence of \MTwoA/\MZerA.

With a larger, more representative sample of galaxies, we confirm the core 
results of Z09 and SS15 that \MTwoB can \MZerB differ substantially for local 
galaxies.  The degree of scatter seen in \Fig{2d-vs-0d} is comparable to that 
in \Fig{1d-vs-0d} for all datasets considered, except $gi$ SEDs; \MOneB scatter 
to a larger degree than \MTwoB in this case.  These findings appear consistent 
with the suggestion that radial variations in colour are the predominant cause 
of differences between spatially-resolved and integrated mass measurements, 
rather than azimuthal or stochastic (\ie pixel-by-pixel) variations.  However, 
a larger sample of \MTwoB is needed to further test this point.  Until such 
time, our recommendation for the use of \MZerA, \MOneA, or \MTwoB depend on the 
particular application.  For accuracies $<$0.1 dex and/or small samples, we 
advocate a spatially-resolved approach (1D/2D), while statistical corrections 
for the use of integrated light may be invoked in all other cases (see also 
SS15).


\section{Conclusions}\label{sec:conc}

Drawing on the literature, SPS models, and a representative sample of local 
galaxies, we have quantified the random + systematic uncertainties in stellar 
mass estimates derived from MLCRs.  In particular, we have searched for 
possible biases in \MStrB due to the use of MLCRs (as opposed to SED fitting), 
prior assumptions invoked in building MLCRs, scope of SED sampling, and 
modelling integrated versus spatially-resolved data.  Both MLCRs and SED 
fitting are able to recover the masses of mock galaxies to within $\sim$0.2 dex 
of scatter (even when limited to just $g-$ and $i-$band photometry), and yield 
equivalent results when applied to real galaxies.  We find that the priors 
dominate the systematics, contributing at most $\sim$0.3 or 0.6 dex of 
uncertainty, depending on whether optical or optical+NIR photometry is used.  
Modelling integrated photometry is the next most important effect.  In so 
doing, 0.06-0.07 dex of scatter is introduced in \MStrB measurements, while
an additional bias of +0.05-0.07 dex results from including NIR data.  The 
latter effect may be an indication that SPS models lack consistency throughout 
the optical-NIR range.  From the MLCR perspective, while the use of a single 
colour (e.g. $g-i$) to constrain \MLB leads to biases on the order of 
$\sim$10\%, increased sampling of galaxy SEDs beyond a few filters spanning the 
optical-NIR range only serves to reduce the statistical errors in the results.

Our investigation demonstrate that the MLCR and SED fitting methods yield
comparable results, as expected since their assessment of \MStrB both depend
on SPS models.  Therefore, improvements to the MLCR method should come
from extensive analyses of resolved stellar populations in galaxies, whereby
fundamental parameters are most accurately constrained via colour-magnitude 
diagrams.  In the meantime, we advocate the use of the MLCR method for
applications focussed on exceptionally large datasets and/or complex parameter 
spaces.  Barring future evidence to the contrary, we only recommend the use of 
a spatially-resolved approach to \MStrB measurements for cases involving either 
small numbers of galaxies or if accuracies to better than 0.1 dex are required.

\bigskip

JCR and SC acknowledge financial support from the National Science and 
Engineering Council of Canada through a post-graduate scholarship and Discovery 
Grant, respectively.  We also thank the referee, Stefano Zibetti, for a very 
thorough and constructive report, and Michael McDonald for providing the AB 
photometric zero-points to calibrate our surface brightness profiles extracted 
from SDSS images.

This research has made use of (i) the NASA/IPAC Extragalactic Data base (NED) 
which is operated by the Jet Propulsion Laboratory, California Institute of 
Technology, under contract with the National Aeronautics and Space 
Administration, as well as NASA's Astrophysics Data System; and (ii) the Sloan 
Digital Sky Survey (SDSS).  Funding for the creation and distribution of the 
SDSS Archive has been provided by the Alfred P. Sloan Foundation, the 
Participating Institutions, the National Aeronautics and Space Administration, 
the National Science Foundation, the U.S. Department of Energy, the Japanese 
Monbukagakusho, and the Max Planck Society.  The SDSS Web site is 
http://www.sdss.org/.  The SDSS is managed by the Astrophysical Research 
Consortium (ARC) for the Participating Institutions.



\clearpage

\appendix

\section{Appendix}\label{sec:app}

\Table{mlcr-fits} provides linear least-squares fits to the MLCRs created on 
the basis of the BC03 and FSPS SPS models (see 
\Figs{comp-mlcr-opt}{comp-mlcr-opt+nir}).  Fits are provided for each of the 
filter and colour combinations accessed by our dataset.  The coefficients in 
this table follow the generic equation: $\log (M_*/L)_{\lambda} = m_{\lambda} 
\times (colour) + b_{\lambda}.$  We omit the CB07 MLCRs as those can be found 
in Table B1 of Z09.  Table 3 shows the Pearson correlation coefficient for each 
of these fits.  In all cases the coefficient is greater than 0.55, and 
$\sim$70\% of the time it is greater than 0.85.

\begin{deluxetable}{lcccccccccc}
 \tabletypesize{\footnotesize}
 \tablewidth{0pc}  
 \tablecaption{Coefficients of linear fits for the BC03 and FSPS MLCRs.}
 \tablehead{
  \colhead{Colour} &
  \colhead{$m_g$} &
  \colhead{$b_g$} &
  \colhead{$m_r$} &
  \colhead{$b_r$} &
  \colhead{$m_i$} &
  \colhead{$b_i$} &
  \colhead{$m_z$} &
  \colhead{$b_z$} &
  \colhead{$m_H$} &
  \colhead{$b_H$} \\
  \colhead{(1)}  &
  \colhead{(2)}  &
  \colhead{(3)}  &
  \colhead{(4)}  &
  \colhead{(5)}  &
  \colhead{(6)}  &
  \colhead{(7)}  &
  \colhead{(8)}  &
  \colhead{(9)}  &
  \colhead{(10)} &
  \colhead{(11)}
 }
 \startdata
  \multicolumn{11}{l}{BC03:} \cr
  $g-r$ & 2.029 & -0.984 & 1.629 & -0.792 & 1.438 & -0.771 & 1.306 & -0.796 & 0.980 & -0.920 \cr
  $g-i$ & 1.379 & -1.067 & 1.110 & -0.861 & 0.979 & -0.831 & 0.886 & -0.848 & 0.656 & -0.950 \cr
  $g-z$ & 1.116 & -1.132 & 0.900 & -0.916 & 0.793 & -0.878 & 0.716 & -0.888 & 0.521 & -0.968 \cr
  $g-H$ & 0.713 & -1.070 & 0.577 & -0.870 & 0.507 & -0.834 & 0.454 & -0.842 & 0.313 & -0.902 \cr
  $r-i$ & 4.107 & -1.170 & 3.325 & -0.952 & 2.925 & -0.908 & 2.634 & -0.912 & 1.892 & -0.977 \cr
  $r-z$ & 2.322 & -1.211 & 1.883 & -0.987 & 1.655 & -0.937 & 1.483 & -0.935 & 1.038 & -0.975 \cr
  $r-H$ & 1.000 & -0.988 & 0.814 & -0.809 & 0.713 & -0.778 & 0.634 & -0.786 & 0.414 & -0.833 \cr
  $i-z$ & 5.164 & -1.212 & 4.201 & -0.991 & 3.683 & -0.939 & 3.283 & -0.931 & 2.210 & -0.947 \cr
  $i-H$ & 1.257 & -0.869 & 1.024 & -0.713 & 0.895 & -0.693 & 0.792 & -0.706 & 0.495 & -0.761 \cr
  $z-H$ & 1.615 & -0.729 & 1.316 & -0.600 & 1.150 & -0.593 & 1.015 & -0.616 & 0.615 & -0.692 \cr
  \cr
  \multicolumn{11}{l}{FSPS:} \cr
  $g-r$ & 1.897 & -0.811 & 1.497 & -0.647 & 1.281 & -0.602 & 1.102 & -0.583 & 0.672 & -0.605 \cr
  $g-i$ & 1.231 & -0.805 & 0.973 & -0.644 & 0.831 & -0.597 & 0.713 & -0.576 & 0.426 & -0.592 \cr
  $g-z$ & 0.942 & -0.764 & 0.744 & -0.612 & 0.634 & -0.568 & 0.542 & -0.548 & 0.316 & -0.565 \cr
  $g-H$ & 0.591 & -0.655 & 0.468 & -0.527 & 0.398 & -0.494 & 0.339 & -0.482 & 0.191 & -0.515 \cr
  $r-i$ & 3.374 & -0.745 & 2.675 & -0.600 & 2.275 & -0.556 & 1.940 & -0.537 & 1.120 & -0.554 \cr
  $r-z$ & 1.795 & -0.670 & 1.421 & -0.539 & 1.206 & -0.502 & 1.021 & -0.487 & 0.570 & -0.513 \cr
  $r-H$ & 0.824 & -0.542 & 0.654 & -0.439 & 0.555 & -0.418 & 0.469 & -0.414 & 0.254 & -0.463 \cr
  $i-z$ & 3.709 & -0.550 & 2.933 & -0.443 & 2.484 & -0.419 & 2.084 & -0.411 & 1.112 & -0.457 \cr
  $i-H$ & 1.073 & -0.460 & 0.852 & -0.375 & 0.722 & -0.362 & 0.608 & -0.365 & 0.322 & -0.430 \cr
  $z-H$ & 1.493 & -0.414 & 1.188 & -0.339 & 1.008 & -0.333 & 0.849 & -0.341 & 0.449 & -0.417 \cr
 \enddata
 \label{tbl:mlcr-fits}
\end{deluxetable}

\begin{deluxetable}{lccccc}
 \tabletypesize{\footnotesize}
 \tablewidth{0pc}
 \tablecaption{Pearson correlation coefficients for the BC03 and FSPS MLCRs.}
 \tablehead{
  \colhead{Colour} &
  \colhead{$r_g$} &
  \colhead{$r_r$} &
  \colhead{$r_i$} &
  \colhead{$r_z$} &
  \colhead{$r_H$} \\
  \colhead{(1)}  &
  \colhead{(2)}  &
  \colhead{(3)}  &
  \colhead{(4)}  &
  \colhead{(5)}  &
  \colhead{(6)}
 }
 \startdata
  \multicolumn{6}{l}{BC03:} \cr
  $g-r$ & 0.98 & 0.97 & 0.96 & 0.96 & 0.92 \cr
  $g-i$ & 0.99 & 0.98 & 0.97 & 0.97 & 0.91 \cr
  $g-z$ & 0.98 & 0.98 & 0.97 & 0.96 & 0.89 \cr
  $g-H$ & 0.95 & 0.95 & 0.94 & 0.93 & 0.82 \cr
  $r-i$ & 0.98 & 0.97 & 0.97 & 0.96 & 0.88 \cr
  $r-z$ & 0.96 & 0.96 & 0.95 & 0.94 & 0.84 \cr
  $r-H$ & 0.90 & 0.90 & 0.89 & 0.87 & 0.72 \cr
  $i-z$ & 0.92 & 0.92 & 0.91 & 0.90 & 0.77 \cr
  $i-H$ & 0.85 & 0.85 & 0.84 & 0.82 & 0.65 \cr
  $z-H$ & 0.81 & 0.82 & 0.80 & 0.78 & 0.60 \cr
  \cr
  \multicolumn{6}{l}{FSPS:} \cr
  $g-r$ & 0.95 & 0.93 & 0.91 & 0.88 & 0.74 \cr
  $g-i$ & 0.96 & 0.93 & 0.91 & 0.88 & 0.73 \cr
  $g-z$ & 0.95 & 0.93 & 0.91 & 0.87 & 0.70 \cr
  $g-H$ & 0.94 & 0.92 & 0.90 & 0.86 & 0.66 \cr
  $r-i$ & 0.94 & 0.92 & 0.90 & 0.86 & 0.69 \cr
  $r-z$ & 0.93 & 0.91 & 0.88 & 0.84 & 0.65 \cr
  $r-H$ & 0.91 & 0.90 & 0.87 & 0.83 & 0.62 \cr
  $i-z$ & 0.90 & 0.88 & 0.85 & 0.80 & 0.59 \cr
  $i-H$ & 0.90 & 0.88 & 0.85 & 0.81 & 0.59 \cr
  $z-H$ & 0.89 & 0.88 & 0.85 & 0.81 & 0.59 \cr
 \enddata
 \label{tbl:mlcr-fits-r}
\end{deluxetable}


\clearpage

\begin{thebibliography}{99}

\bibitem[Abazajian et al.(2009)]{A09} Abazajian, K.~N., Adelman-McCarthy, 
 J.~K., Ag{\"u}eros, M.~A., et al.\ 2009, \apjs, 182, 543
\bibitem[Behroozi et al.(2010)]{B10} Behroozi, P.~S., Conroy, C., \& Wechsler, 
 R.~H.\ 2010, \apj, 717, 379
\bibitem[Bell \& de Jong(2001)]{BdJ01} Bell, E.~F., \& de Jong, R.~S.\ 2001, 
 \apj, 550, 212
\bibitem[Bell et al.(2003)]{B03} Bell, E.~F., McIntosh, D.~H., Katz, N., \& 
 Weinberg, M.~D.\ 2003, \apjs, 149, 289
\bibitem[Bruzual \& Charlot(2003)]{BC03} Bruzual, G., \& Charlot, S.\ 2003, 
 MNRAS, 344, 1000
\bibitem[Bruzual(2007)]{B07} Bruzual, A.~G.,\ 2007, in Vazdekis A.,
  Peletier R. F., eds, Proc. IAU Symp. 241, Stellar Populations as
  Building Blocks of Galaxies. Cambridge University Press, Cambridge,
  p. 125
\bibitem[Calzetti et al.(2000)]{C00} Calzetti, D., Armus, L., Bohlin, R.~C., et 
 al.\ 2000, \apj, 533, 682
\bibitem[Cappellari \& Copin(2003)]{CC03} Cappellari, M., \& Copin, Y.\ 2003, 
 \mnras, 342, 345
\bibitem[Chabrier(2003)]{C03} Chabrier, G.\ 2003, \pasp, 115, 763
\bibitem[Charlot \& Fall(2000)]{CF00} Charlot, S., \& Fall, S.~M.\ 2000, \apj, 
 539, 718
\bibitem[Conroy et al.(2009)]{C09} Conroy, C., Gunn, J.~E., \& White, M.\ 2009, 
 \apj, 699, 486
\bibitem[Conroy(2013)]{C13} Conroy, C.\ 2013, \araa, 51, 393
\bibitem[Courteau(1996)]{C96} Courteau, S.\ 1996, \apjs, 103, 363
\bibitem[Courteau et al.(2014)]{C14} Courteau, S., Cappellari, M., de Jong, 
 R.~S., et al.\ 2014, Reviews of Modern Physics, 86, 47
\bibitem[da Cunha et al.(2008)]{dC08} da Cunha, E., Charlot, S., \& Elbaz, D.\ 
 2008, \mnras, 388, 1595
\bibitem[Gallazzi \& Bell(2009)]{GB09} Gallazzi, A., \& Bell, E.~F.\ 2009, 
 \apjs, 185, 253
\bibitem[Gallazzi et al.(2005)]{G05} Gallazzi, A., Charlot, S., Brinchmann, J., 
 White, S.~D.~M., \& Tremonti, C.~A.\ 2005, \mnras, 362, 41
\bibitem[Gavazzi et al.(2003)]{G03} Gavazzi, G., Boselli, A., Donati, A., 
 Franzetti, P., \& Scodeggio, M.\ 2003, A\&A, 400, 451
\bibitem[Hall et al.(2012)]{H12} Hall, M., Courteau, S., Dutton, A.~A., 
 McDonald, M., \& Zhu, Y.\ 2012, \mnras, 425, 2741
\bibitem[Into \& Portinari(2013)]{IP13} Into, T., \& Portinari, L.\ 2013, 
 \mnras, 430, 2715
\bibitem[Longhetti \& Saracco(2009)]{LS09} Longhetti, M., \& Saracco, P.\ 2009, 
 \mnras, 394, 774
\bibitem[Mannucci et al.(2010)]{M10} Mannucci, F., Cresci, G., Maiolino, R., 
 Marconi, A., \& Gnerucci, A.\ 2010, \mnras, 408, 2115
\bibitem[Maraston(2005)]{M05} Maraston, C.\ 2005, \mnras, 362, 799
\bibitem[Marigo \& Girardi(2007)]{MG07} Marigo, P., \& Girardi, L.\ 2007, \aap, 
 469, 239
\bibitem[Marigo et al.(2008)]{M08} Marigo, P., Girardi, L., Bressan, A., et 
 al.\ 2008, \aap, 482, 883
\bibitem[Mei et al.(2007)]{M07} Mei, S., et al.\ 2007, ApJ, 655, 144
\bibitem[McDonald et al.(2009)]{M09} McDonald, M., Courteau, S., \& Tully, 
 R.~B.\ 2009, MNRAS, 394, 2022
\bibitem[McDonald et al.(2011)]{M11} McDonald, M., Courteau, S., Tully, R.~B., 
 \& Roediger, J.\ 2011, \mnras, 414, 2055
\bibitem[McGaugh \& Schombert(2014)]{MS14} McGaugh, S.~S., \& Schombert, J.~M.\ 
 2014, \aj, 148, 77
\bibitem[Roediger et al.(2011a)]{R11a} Roediger, J.~C., Courteau, S., McDonald, 
 M., \& MacArthur, L.~A.\ 2011, \mnras, 416, 1983
\bibitem[Schlafly \& Finkbeiner(2011)]{SF11} Schlafly, E.~F., \& Finkbeiner, 
 D.~P.\ 2011, \apj, 737, 103
\bibitem[Schlegel et al.(1998)]{S98} Schlegel, D.~J., Finkbeiner, D.~P., \& 
 Davis, M.\ 1998, \apj, 500, 525
\bibitem[Skrutskie et al.(2006)]{S06} Skrutskie, M.~F., et al.\ 2006, AJ, 131, 
 1163
\bibitem[Sorba \& Sawicki(2015)]{SS15} Sorba, R., \& Sawicki, M.\  2015, 
 arXiv:1506.01653
\bibitem[Taylor et al.(2011)]{T11} Taylor, E.~N., Hopkins, A.~M., Baldry, 
 I.~K., et al.\ 2011, \mnras, 418, 1587
\bibitem[Tremonti et al.(2004)]{T04} Tremonti, C.~A., Heckman, T.~M., 
 Kauffmann, G., et al.\ 2004, \apj, 613, 898
\bibitem[van Albada \& Sancisi(1986)]{vAS86} van Albada, T.~S., \& Sancisi, R.\ 
 1986, Royal Society of London Philosophical Transactions Series A, 320, 447
\bibitem[van Dokkum(2008)]{vD08} van Dokkum, P.~G.\ 2008, \apj, 674, 29
\bibitem[Zibetti(2009)]{Z09a} Zibetti, S.\ 2009, arXiv:0911.4956
\bibitem[Zibetti et al.(2009)]{Z09b} Zibetti, S., Charlot, S., \& Rix, H.-W.\ 
 2009, \mnras, 400, 1181

\end{thebibliography}
\end{document}